\documentclass[lettersize,journal]{IEEEtran}

\usepackage{graphicx,psfrag,epsfig,epsf,latexsym,hhline,amsmath,amssymb,multirow}
\usepackage{pst-plot}
\usepackage{pstricks-add}
\usepackage{pifont}
\usepackage{graphicx}
\usepackage{amsthm}
\usepackage[linesnumbered,ruled,vlined]{algorithm2e}
\usepackage[noadjust]{cite}
\usepackage{array}
\usepackage{blindtext}
\usepackage{etoolbox}
\usepackage{comment}
\usepackage{epstopdf}
\usepackage{hyperref}

\SetCommentSty{mycommfont}

\SetKwInput{KwInput}{Input}                
\SetKwInput{KwOutput}{Output}              

\newcolumntype{M}[1]{>{\centering\arraybackslash}m{#1}}
\newcolumntype{P}[1]{>{\centering\arraybackslash}p{#1}}

\newcommand\xqed[1]{%
  \leavevmode\unskip\penalty9999 \hbox{}\nobreak\hfill
  \quad\hbox{#1}}
\newcommand\demo{\xqed{$\blacksquare$}}

\newcommand{\msf}[1]{\mathsf{#1}}

\newcommand{\SNR}{\msf{SNR}}

\newcommand{\Pp}{\mathbb{P}}
\newcommand{\E}{\mathbb{E}}

\newcommand{\iid}{i.\@i.\@d.\ }

\theoremstyle{definition}
\newtheorem{definition}{Definition}
\newtheorem{example}{Example}

\newtheorem{lemma}{Lemma}
\newtheorem{proposition}{Proposition}
\newtheorem{remark}{Remark}

\begin{document}
\title{Downlink Transmission with Heterogeneous URLLC Services: Discrete Signaling With Single-User Decoding}
\author{Min Qiu,~\IEEEmembership{Member,~IEEE},
Yu-Chih Huang,~\IEEEmembership{Member,~IEEE}, and Jinhong Yuan,~\IEEEmembership{Fellow,~IEEE}

\thanks{The work of Min Qiu and Jinhong Yuan was supported in part by the Australian Research Council (ARC) Discovery Project under Grant DP220103596, and in part by the ARC Linkage Project under Grant LP200301482. The work of Yu-Chih Huang was supported by the National Science and Technology Council, Taiwan, under Grant NSTC 111-3114-E-A49-001-. This work was also supported in part by the Higher Education Sprout Project of the National Yang Ming Chiao Tung University and Ministry of Education (MOE), Taiwan. \emph{(Corresponding author: Yu-Chih Huang)}}

\thanks{
This work will be presented in part at the 2023 IEEE International Conference on Communications (ICC), Rome, Italy \cite{ICCQiu23}.

Min Qiu and Jinhong Yuan are with the School of Electrical Engineering and Telecommunications, University of New South Wales, Sydney, NSW, 2052 Australia (e-mail: min.qiu@unsw.edu.au; j.yuan@unsw.edu.au).
Yu-Chih Huang is with the Institute of Communications Engineering, National Yang Ming Chiao Tung University, Hsinchu 300, Taiwan (e-mail: jerryhuang@nycu.edu.tw).
}%
}

\maketitle

\begin{abstract}
The problem of designing downlink transmission schemes for supporting heterogeneous ultra-reliable low-latency communications (URLLC) and/or with other types of services is investigated. We consider the broadcast channel, where the base station sends superimposed signals to multiple users. Under heterogeneous blocklength constraints, strong users who are URLLC users cannot wait to receive the entire transmission frame and perform successive interference cancellation (SIC) due to stringent latency requirements, in contrast to the conventional infinite blocklength cases. Even if SIC is feasible, SIC may be imperfect under finite blocklength constraints. To cope with the heterogeneity in latency and reliability requirements, we propose a practical downlink transmission scheme with \emph{discrete signaling} and \emph{single-user decoding (SUD), i.e., without SIC}. We carefully design the discrete input distributions to enable efficient SUD by exploiting the structural interference. Furthermore, we derive the second-order achievable rate under heterogenous blocklength and error probability constraints and use it to guide the design of channel coding and modulations. It is shown that in terms of achievable rate under short blocklength, the proposed scheme with regular quadrature amplitude modulations and SUD can operate \emph{extremely close} to the benchmark schemes that assume perfect SIC with Gaussian signaling.

\end{abstract}


\section{Introduction}\label{sec:intro}
Ultra-reliable low-latency communication (URLLC) is one of the most important usage scenarios in the fifth generation (5G) communication systems and beyond. It accommodates applications and services with stringent latency and reliability requirements, such as industrial automation and remote surgery. In particular, URLLC is required to achieve at least $99.9999\%$ reliability with 1 \emph{ms} end-to-end latency \cite{TR38.913}, which poses a great challenge to communication system design. Significant efforts have been made for trying to achieve these two goals from the perspectives of radio resource management, signal processing, and channel coding \cite{7945856,8469808,8594709}.

In 5G communication systems, heterogeneous services such as URLLC and enhanced mobile broadband (eMBB), are allowed to coexist within the same network architecture by network slicing \cite{8004168}, which allocates orthogonal resources to heterogeneous services to guarantee their mutual isolation. Such an orthogonal approach is difficult to accommodate the growing number of devices and services due to its inefficient use of resources. Academic research and industries have been actively investigating effective coexistence mechanisms of heterogeneous services for achieving higher spectrum and energy efficiency with per-service guarantees \cite{TR38.802,8403963,8476595,8647460,21p915,9562192,9831059}. One may refer the problem of designing coexistence schemes back to the classical studies of the broadcast channel (BC) and the multiple access channel (MAC) \cite{Cover:2006:EIT:1146355}. Specifically, for the scalar downlink BC where the base station serves multiple users simultaneously, it is well known that superposition coding and successive interference cancellation (SIC) are two key ingredients for achieving the whole capacity region effectively \cite{tse_book}. In contrast to the conventional orthogonal multiple access, many popular multiple access schemes have adopted these two techniques for enabling simultaneous communications between multiple users and the base station in the same time-frequency resources \cite{Ding17J,8876877,9693417,9831440,RSMA_JSAC}

Despite these results, the challenges from the coexistence between heterogeneous URLLC services and other services have not been fully addressed. First, URLLC services cannot leverage SIC decoding \cite{8403963,8476595,9831059}. This is because SIC introduces additional delay and complexity, and may affect the reliability of the URLLC services. To see this, consider the downlink BC where the base station performs superposition coding and sends the superposition of URLLC and eMBB symbol blocks to all users. Performing SIC in the traditional sense at the URLLC device requires the reception of the whole superimposed symbol block up to the length of the longest eMBB symbol block, causing significant reception delay. Since SIC requires decoding of interfering codewords before decoding the desired codeword, it also introduces extra decoding delay and complexity. Note also that in the downlink BC, SIC is performed at the device side which usually has limited power and computational resources. Thus, the complexity of SIC is also a concern. In addition, because of the heterogeneous reliability requirements, decoding eMBB then URLLC in a SIC fashion would require the decoding of eMBB services with at least the same reliability as for URLLC services. This may not be achievable since eMBB messages are coded based on a lower reliability requirement compared to URLLC services. As a result, SIC at URLLC devices is likely to fail, which can introduce error propagation during the decoding of URLLC codewords and affect the reliability. Second, the characterization of the achievable rate from most works on multiple access schemes and URLLC-eMBB coexistence schemes is based on the infinite blocklength assumption. However, these results may not precisely describe the performance behavior of URLLC services which have short blocklength packets \cite{5452208,7529226}. Some works such as \cite{8277977,8345745,8933345} attempt to refine the performance analysis by using the point-to-point additive white Gaussian noise (AWGN) finite blocklength achievable rate \cite{5452208} and assume homogeneous blocklength and error probability constraints. Whereas, heterogeneous services such as URLLC and eMBB have their own blocklength and error probability requirements. Since symbol blocks with different lengths are superimposed and sent to all users, the received symbol block suffers from heterogeneous interference statistics across symbols.

To reduce the SIC reception latency, \cite{8909370} uses two short codewords to form a long eMBB symbol block such that SIC can be performed once the first short codeword is received. Alternatively, \cite{9518265,9838392} leverage early decoding \cite{6875095} to decode the partially received interfering codeword for SIC only when certain channel and blocklength conditions are met. However, generalizing the schemes in \cite{8909370,9518265,9838392} to more than two users is very difficult. Yet, the decoding delay and complexity introduced by SIC increase with the number of users and could not be eliminated. In contrast, treating interference as noise (TIN) is commonly adopted in most practical communication systems as it simply involves single-user decoding (SUD). Compared to SIC, TIN achieves strictly lower latency and complexity and does not have error propagation. It is also worth noting that all the aforementioned works assume capacity-achieving input distributions such as Gaussian signaling or shell codes, i.e., codewords drawn from a power shell \cite[Sec. X]{6767457}. Apart from the implementation difficulties in practical systems, these signalings have worse performance when using TIN compared to using time-sharing \cite{8476595}. In practical communication systems, the current prevailing approach is to adopt channel coding with discrete constellations, e.g., quadrature amplitude modulation (QAM) \cite{TS138212_v16p8}. Motivated by the benefits and challenges of the coexistence between heterogeneous services in downlink, we study the downlink BC and design new coexistence scheme for simultaneously serving different types of URLLC services and/or other heterogeneous services, e.g., eMBB, based on practical coding, modulations, and TIN, aiming to achieve rates close to capacity-achieving signaling with perfect SIC while satisfying per-service requirements.

Discrete signaling and TIN for the Gaussian BC and the Gaussian interference channel in the infinite blocklength regime have been investigated in our previous works \cite{8291591,9535131}. However, under \emph{heterogeneous finite blocklength} and \emph{non-vanishing error probability} constraints, how to effectively manage heterogeneous interference across received symbol sequences has not been investigated. Moreover, the characterization of second-order achievable rates with practical coded modulations and TIN for this scenario is lacking. In this paper, we provide a fundamental study on the behavior of practical coded modulation systems designed for heterogeneous URLLC services. The main contributions are as follows.
\begin{itemize}
\item We consider a $K$-user downlink BC model with heterogeneous blocklength and error probability constraints. For this model, we first divide the intended symbol block for each user into sub-blocks according to the signal power and interference statistics. We then design the modulation and power for each sub-block to exploit the heterogeneous and structural interference across the superimposed symbols. Note that unlike \cite{8909370}, in our scheme each user employs only one channel code while each sub-block adopts its own constellation and sub-block power, which may be different from others. To keep the encoding and decoding complexity as low as in the single-user case, each use only uses a single channel code and performs TIN decoding.

\item We derive the second-order achievable rate of each user under TIN for the $K$-user BC for a given individual discrete constellation, power assignment, blocklength, and target error probability. Our approach involves the use of dependence testing bound \cite[Th. 17]{5452208} for upper bounding the TIN decoding error probability and employing the Berry-Esseen central limit theorem \cite[Th. 2, Ch. XVI-5]{Feller_book} to obtain the second-order terms in the achievable rate. The impacts of the interfering symbols with heterogeneous lengths can be revealed through the derived achievable rate. With the derived achievable rate, we show that the problem of designing modulations and codes for the proposed scheme can be transformed into that for the point-to-point case, which significantly simplifies the problem.

\item We first provide the second-order achievable rate simulation for the proposed scheme and compare them with the benchmark schemes that assume Gaussian and shell codes with perfect SIC. Interestingly, it is shown that the channel dispersion of the proposed QAM signaling with TIN is smaller than those of Gaussian and shell codes, respectively, with perfect SIC. Since short blocklength and ultra-low target error probability are the main features of URLLC communication scenarios, the second-order term has a substantial impact on the achievable rate. Thus, under these scenarios, the proposed scheme with QAM and TIN can achieve rate pairs very close to those of the benchmark schemes with perfect SIC. The effectiveness of the proposed scheme is further demonstrated via the error performance of a practical set-up of our scheme, where off-the-shelf channel codes are employed.

\end{itemize}

\section{System Model}\label{sec:model}

\begin{figure}[t!]
	\centering
\includegraphics[width=\linewidth]{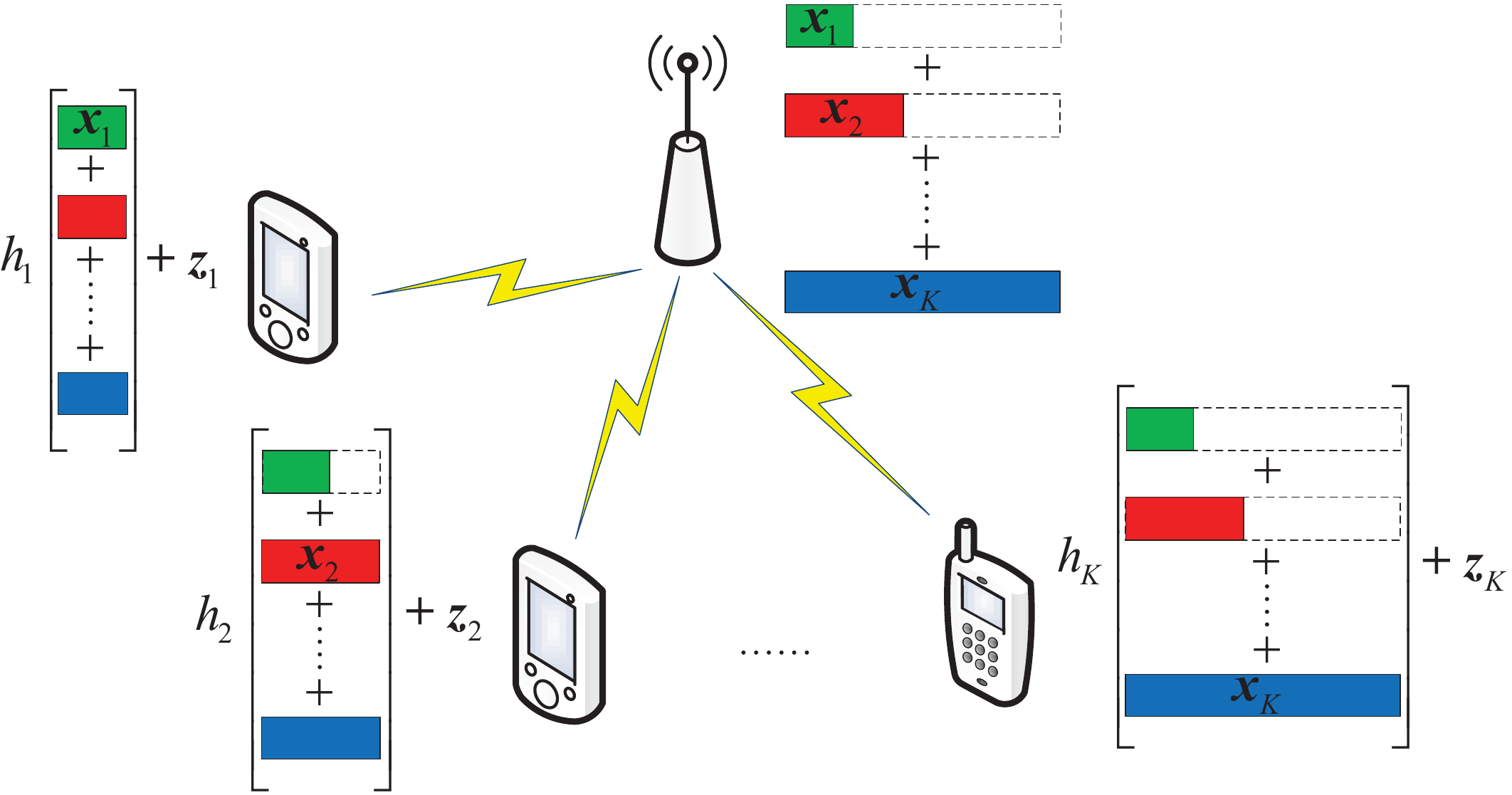}
\caption{System model of the $K$-user downlink BC under heterogeneous blocklength constraints.}
\label{fig:model}
\end{figure}

We consider a scalar downlink BC that consists of one transmitter and $K$ receivers\footnote{For possible extensions to multiple antenna systems, one may use zero-forcing beamforming to null out interference and convert the problem into multiple single antenna problems as in \cite{8108239}.}. The system model is depicted in Fig. \ref{fig:model}. We denote by $\boldsymbol{x}_k \in \mathbb{C}^{N_k}$ the transmitted packet of coded symbols after power allocation for user $k$, where $k\in \{1,\ldots,K\}$ and $N_k$ denotes the symbol length. We assume that the symbol lengths satisfy $N_1 \leq N_2 \leq,\ldots,\leq N_K$ without loss of generality. Note that the $K$ users can be a combination of different types of URLLC users and/or with other types of users, e.g., eMBB users. The transmitter broadcasts the superimposed coded symbols $\boldsymbol{x}$ of length $N_K$ to $K$ users, where
\begin{align}\label{eq:sys1}
\boldsymbol{x} =& ([\boldsymbol{x}_1,\boldsymbol{0}^{N_K-N_1}]+[\boldsymbol{x}_2,\boldsymbol{0}^{N_K-N_2}]+\ldots+\boldsymbol{x}_K) \nonumber \\
=&\sum\nolimits^K_{k=1}\boldsymbol{x}'_k \in \mathbb{C}^{N_K}.
\end{align}
For user $k$, $\boldsymbol{0}^{N_K-N_k}$ represent the zero padding\footnote{For ease of presentation, we use the term zero padding to address the differences among blocklengths. Operationally, no zero padding is required and the transmitter simply superimposes the signals of different lengths.} of length $N_K-N_k$. We define $\boldsymbol{x}'_k \triangleq [\boldsymbol{x}_k,\boldsymbol{0}^{N_K-N_k}]$ to be the vector which contains the transmitted packet and zero padding. User $K$ has the longest packet length and does not require zero padding. Due to the urgency of URLLC applications, packet $\boldsymbol{x}_k$ needs to be transmitted as soon as possible. Thus, zero padding is placed after each packet.
We introduce the following individual power constraint $P_k$ and total power constraint $P$, and with $P_k\leq P$.
\begin{align}
&\frac{1}{N_k}\sum\nolimits_{j=1}^{N_k}|x_k[j]|^2 \leq P_k,\label{eq:ind_power} \\
&\frac{1}{N_K}\sum\nolimits^{N_K}_{j=1}(|x'_1[j] |^2+\ldots+|x'_K[j] |^2) \leq P.\label{eq:total_power}
\end{align}

We denote by $h_k \in \mathbb{C}$ the channel of user $k$, $k\in\{1,\ldots,K\}$. We assume that $h_k$ is subject to quasi-static fading, i.e., $h_k$ remains unchanged over the duration of $N_K$ symbol periods. As illustrated in Fig. \ref{fig:model}, the received signal at user $k$ is given by
\begin{align}\label{eq:yk}
y_k[j] =& h_kx[j]+z_k[j]\nonumber\\
=& \left\{ {\begin{array}{l}
h_k\sum^K_{k'=1}x_{k'}[j]+z_k[j],j=1,\ldots,N_1\\
h_k\sum^K_{k'=2}x_{k'}[j]+z_k[j], j=N_1+1,\ldots,N_2\\
\vdots \\
h_k\sum^K_{k'=k}x_{k'}[j]+z_k[j], j=N_{k-1}+1,\ldots,N_k\\
\end{array}} \right.,
\end{align}
where $z_k[j] \sim \mathcal{CN}(0,1)$ is the i.i.d. Gaussian noise. Due to the latency constraint, user $k$ starts decoding once it receives the first $N_k$ symbols of $\boldsymbol{y}_k$. One can see that except for user 1, the transmitted packet of other users experiences \emph{heterogeneous} interference strength across its symbols. This is significantly different from the homogeneous blocklength case where each symbol suffers from the same level of interference strength. From \eqref{eq:sys1} and \eqref{eq:yk}, we see that $\boldsymbol{x}_k$ can be divided into $k$ sub-blocks, where for $k'\in\{1,\ldots,k\}$, the $k'$-th sub-block will be interfered by the symbol blocks of users with index set $\{k',\ldots,K\}\setminus k$. We note that the position of zero padding in $\boldsymbol{x}$ can affect the interference distribution on each user's intended symbol sequence. If the transmission delay is not a concern, it is possible to design the zero padding position for each user to maximize the overall achievable rate region. However, this task is beyond the scope of the paper. We emphasize that zero padding positions do not affect our design principle and analysis as we consider low-complexity TIN decoding. In contrast, this will have a significant impact on the design and analysis of the SIC based schemes, e.g., \cite{8909370,9518265,9838392}.

We assume that the transmitter has the knowledge of channel magnitudes while each receiver has full channel state information about its channel. We stress that even for this fundamental channel model, many problems are yet to be solved, e.g., the optimal communication strategy, the optimal input distribution, and the second-order converse. Two important performance metrics are jointly considered in this paper, namely the achievable rate $R_k$ and upper bound of the average decoding error probability $\epsilon_k$ for user $k\in\{1,\ldots,K\}$.

\section{Preliminaries}\label{sec:preliminaries}
In this section, we briefly review the definitions for QAM, Gaussian codes, shell codes, TIN decoding, and SIC decoding from \cite{GamalKim11,7056434,7605463}. All logarithms are base 2. $Q^{-1}(x)$ denotes the inverse of Q function $Q(x)=\int^{\infty}_{x}\frac{1}{\sqrt{2\pi}}e^{-\frac{t^2}{2}}dt$. We write $f(x) = O(g(x))$ if $\exists M\in\mathbb{R}^+,x_0\in\mathbb{R}$ such that $|f(x)| \leq Mg(x), \forall x\geq x_0$. Random variables are represented by uppercase letters, e.g., $X$, and their realizations are represented by lowercase letters, e.g., $x$. $X^{[n]}$ denotes the sequence $X[1],\ldots,X[n]$.

A regular QAM $\mathcal{X}$ with zero mean and minimum distance $d_{\min}(\mathcal{X})>0$ has average energy $\E_{X}[|X|^2]=\frac{1}{|\mathcal{X}|}\sum\limits_{x\in\mathcal{X}}|x|^2=\frac{|\mathcal{X}|-1}{6}d^2_{\min}(\mathcal{X})$.

\begin{definition}[Gaussian codes \cite{7605463}]\label{def:gauss}
A length-$n$ i.i.d. \emph{Gaussian code} $\boldsymbol{x}$ is defined such that each element of $\boldsymbol{x}$ is distributed over a normal distribution with zero mean and variance $P$, i.e., $\boldsymbol{x} \sim\prod^n_{i=1}\frac{1}{\sqrt{2\pi P}}e^{-\frac{x^2_i}{2P}}$. On the AWGN channel with noise$\sim\mathcal{N}(0,1)$, i.i.d. Gaussian codes achieve a second-order rate of $R_{\text{G}}(P,n,\epsilon) \approx \frac{1}{2}\log(1+P)-\sqrt{\frac{V_\text{G}(P)}{n}}Q^{-1}(\epsilon)$, where $\epsilon$ is the upper bound of the average decoding error probability, and $V_\text{G}(P)=\log^2e\frac{P}{P+1}$ is the Gaussian dispersion.
\demo
\end{definition}

\begin{definition}[Shell codes \cite{7605463}]\label{def:shell}
A length-$n$ \emph{shell code} $\boldsymbol{x}$ is defined such that $\boldsymbol{x}$ is uniformly distributed over a power shell with radius $\sqrt{nP}$, i.e., $\boldsymbol{x} \sim\frac{\delta(\|\boldsymbol{x} \|-\sqrt{nP})}{S_n(\sqrt{nP})}$, where $\delta(.)$ denotes the Dirac delta function, $S_n(r)=\frac{2\pi^{n/2}r^{n-1}}{\Gamma(n/2)}$ is the surface area of an $n$-dimensional sphere with radius $r$ and $\Gamma(.)$ is the Gamma function. On the AWGN channel with noise$\sim\mathcal{N}(0,1)$, shell codes achieve a second-order rate of $R_{\text{S}}(P,n,\epsilon) \approx \frac{1}{2}\log(1+P)-\sqrt{\frac{V_\text{S}(P)}{n}}Q^{-1}(\epsilon)$, where $V_{\text{S}}(P) = \log^2e\frac{P(P+2)}{2(P+1)^2}$ is the shell dispersion.
\demo
\end{definition}

Although both Gaussian codes and shell codes achieve the AWGN channel capacity when $n\rightarrow \infty$, shell codes have been proved to be both \emph{second- and third-order optimal} while Gaussian codes are not \cite{7056434}. One can also notice that since $V_{\text{S}}(P)<V_{\text{G}}(P)$, thus $R_{\text{S}}(P,n,\epsilon)>R_{\text{G}}(P,n,\epsilon)$. Hence, many works in the literature, e.g., \cite{8345745,8933345}, implicitly assume shell codes to obtain higher second-order achievable rates. Note that for the complex channel, both the capacity and dispersion for Gaussian codes and shell codes are twice those in the real channel.

\begin{definition}[Treating interference as noise (TIN) decoding \cite{GamalKim11}]
The decoder performs SUD, e.g., maximum-likelihood (ML) decoding or belief propagation decoding, by treating the interfering codewords as noise. Only the statistical properties such as the signal distribution and power, rather than the actual codebook information, of the interfering codewords, are used.
\demo
\end{definition}

\begin{definition}[Successive interference cancellation (SIC) decoding \cite{GamalKim11}]
The decoder performs single-user decoding to decode the interfering codewords by following some decoding order while treating undecoded codewords as noise. The successfully decoded codeword at each stage is subtracted to remove the interference it causes before moving on to decode the next interfering codeword and ultimately the desired codeword. Full codebook information is required for the decoding of each interfering codeword.
\demo
\end{definition}

TIN has single-user decoding latency and complexity whereas the decoding latency and complexity of SIC grow with the number of users. When Gaussian or shell codes are used, the achievable rate under TIN can be worse than that under time-sharing \cite{8476595}. This can be seen by noting that these capacity-achieving input distributions for the Gaussian channel in fact generate the worst noise (or interference when it is treated as noise) for such channel \cite{Cover:2006:EIT:1146355}. However, discrete signaling can behave differently when being treated as noise, which we are going to demonstrate in this work.

\section{Proposed Discrete Signaling with Single-User Decoding}\label{sec:scheme}
In this section, we introduce the proposed scheme with discrete signaling and TIN decoding. Most importantly, our design takes into account the fact that the received sequences suffer from heterogeneous interference statistics across symbols. Although we consider binary codes and QAM as the underlying channel codes and constellations, respectively, our scheme does not preclude the use of non-binary codes \cite{8066336} and multi-dimensional constellations \cite{8291591}.

\subsection{Two-User Case}\label{sec:2user}
For ease of understanding, we begin by presenting the proposed scheme for the two-user case. We denote the user index by $i\in\{1,2\}$. Recall that in Section \ref{sec:model} we have assumed $N_1 \leq N_2$. Moreover, we further assume that $|h_1|>|h_2|$. This corresponds to the interesting case where the URLLC user, i.e., user 1, is the strong user but performing SIC may not be feasible based on partially received superimposed symbols. The case of $|h_1|<|h_2|$ will be discussed later.

\subsubsection{Encoding}
A length-$k_i$ binary source sequence $\boldsymbol{u}_i$ is encoded into a length-$n_i$ binary codeword $\boldsymbol{c}_i$. Codeword $\boldsymbol{c}_i$ is then interleaved becoming $\boldsymbol{\tilde{c}}_i$ and modulated onto a length-$N_i$ symbol sequence $\boldsymbol{v}_i$, i.e., we use bit-interleaved coded modulations (BICM) \cite{CIT-019}. We emphasize that each user only uses a \emph{single} channel code. This ensures that the encoding and decoding (TIN) complexities for each user are the same as in the single-user case.

\subsubsection{Modulation Mapping}\label{sec:2u_const}
Since $N_1 \leq N_2$, user 2's packet will be \emph{partially interfered}. To handle such heterogeneous interference, we allow user 2 to use two sets of constellations $\Lambda_{2,1}$, and $\Lambda_{2,2}$ while user 1 only needs one constellation set $\Lambda_{1}$. Specifically, the modulated symbols for user 1 satisfy $v_1[j] \in \Lambda_1, \forall j \in \{1,\ldots,N_1\}$, and for user 2 satisfy $v_2[j] \in \Lambda_{2,1}, \forall j \in \{1,\ldots,N_1\}$, and $v_2[j] \in \Lambda_{2,2}, \forall j \in \{N_1+1\ldots,N_2\}$. In this way, user 2's modulated symbol block $\boldsymbol{v}_2$ can be divided into two sub-blocks, where the first sub-block is interfered by user 1's signals while the second sub-block is interference-free.

We consider $\Lambda_1$, $\Lambda_{2,1}$, and $\Lambda_{2,2}$ to be three regular Gray labeled QAM constellations with zero means and minimum distance 1, and modulation orders $m_1= \log|\Lambda_1|$, $m_{2,1}= \log|\Lambda_{2,1}|$ and $m_{2,2}= \log|\Lambda_{2,2}|$, respectively. For user 1, every $m_1$ bits of the interleaved codeword $\boldsymbol{\tilde{c}}_1$ is mapped to a constellation point of $\Lambda_1$. For user 2, every $m_{2,1}$ bits for the first $N_1m_{2,1}$ bits of $\boldsymbol{\tilde{c}}_2$ is mapped to a constellation point of $\Lambda_{2,1}$. Moreover, every $m_{2,2}$ bits of the last $(N_2-N_1)m_{2,2}$ bits of $\boldsymbol{\tilde{c}}_2$ is mapped to a constellation point of $\Lambda_{2,2}$. The relationship between codeword length $n_i$ and modulated symbol length $N_i$ for user $i$ satisfies
\begin{align}
n_1 =& N_1m_1, \label{u1n1} \\
n_2 =& N_1m_{2,1}+ (N_2-N_1)m_{2,2}.\label{u2n2}
\end{align}

We then introduce the design criteria for modulation orders. To do so, we first need to introduce the following sub-block power constraints. As from above, $\boldsymbol{v}_2$ can be decomposed into two sub-blocks based on the modulated symbols, where the first sub-block will be superimposed to $\boldsymbol{v}_1$. Hence, $\boldsymbol{x}_2$, the symbol block which is after applying power allocation on $\boldsymbol{v}_2$, can also be decomposed into two sub-blocks the same way as for $\boldsymbol{v}_2$. The power constraints for these two sub-blocks of $\boldsymbol{x}_2$ satisfy
\begin{align}\label{eq:sub_power_u2}
\frac{1}{N_i-N_{i-1}}\sum_{j=N_{i-1}+1}^{N_i}|x_2[j]|^2 \leq& P_{2,i}, \;i=1,2,
\end{align}
where we set $N_0=0$. In this way, we can design different power allocations across the first $N_1$ symbols and the last $N_2-N_1$ symbols of $\boldsymbol{x}_2$. With \eqref{eq:ind_power} and \eqref{eq:sub_power_u2}, we establish the following relationship between sub-block power constraints and individual power constraint for user 2 as
\begin{align}
\frac{1}{N_2}\sum_{j=1}^{N_2}|x_2[j]|^2 =& \frac{1}{N_2}\left(\sum_{j=1}^{N_1}|x_2[j]|^2+\sum_{j=N_1+1}^{N_2}|x_2[j]|^2\right) \nonumber \\
\leq& \frac{N_1}{N_2}P_{2,1}+\frac{N_2-N_1}{N_2}P_{2,2}  = P_2.
\end{align}
Additionally, by \eqref{eq:total_power}, the relationship between each sub-block power constraint and the total power constraint satisfies
\begin{align}\label{eq:2userP}
\frac{1}{N_2}&\left(\sum^{N_1}_{j=1}(|x_1[j]|^2+|x_2[j]|^2)+\sum^{N_2}_{j=N_1+1}|x_2[j]|^2\right) \nonumber \\
&\leq \frac{N_1}{N_2}(P_1+P_{2,1})+\frac{N_2-N_1}{N_2}P_{2,2}= P.
\end{align}
With the above sub-block power constraints, we further introduce the following constraints on modulation orders $m_1$, $m_{2,1}$, and $m_{2,2}$
\begin{align}
m_1+m_{2,1} \leq& \left\lfloor\log \left(1+6(P_1+P_{2,1})\max\{|h_1|^2,|h_2|^2\}\right)\right\rfloor,\label{con1}\\
m_{2,1}\leq& \left\lfloor\log \left(6(P_1+P_{2,1})|h_2|^2\right)\right\rfloor,\label{con2}\\
m_{2,2}\leq& \left\lfloor\log \left(1+6P_{2,2}|h_2|^2\right)\right\rfloor,\label{con3}
\end{align}
where the flooring operation $\lfloor .\rfloor$ applies because the modulation orders must be integers. The motivation for introducing the modulation order constraints is to strike a balance between the achievable rate and the interference statistics. This can be seen by noting that under TIN and fixed channel gains, increasing the modulation order of only one user increases its achievable rate (until it reaches its capacity limit) but also introduces more interference to other users. Observe that the RHS of \eqref{con1} is reminiscent of the single-user capacity of the strong user while the RHS of \eqref{con2} is reminiscent of the single-user capacity of user 2. The reason for excluding 1 inside the logarithm of \eqref{con2} while including 6 inside all logarithms is closely related to the minimum distance of individual constellation, which will be explained in Section \ref{sec:2up}c. By inspecting \eqref{con1}, it is also worth noting that the sum capacity of the $K$-user downlink BC can be upper bounded by the single-user capacity of the strongest user \cite[Ch. 6.2.2]{tse_book}. Thus, one can regard \eqref{con1} as a sum-rate constraint. Under the given constraints, we choose the modulation orders such that their sum is close to the sum capacity, where the modulation order alone is the maximum transmission rate by using rate 1 channel codes. Once the modulation orders are determined, the transmission rates can be adjusted by varying the code rates. In addition, \eqref{con1} with $m_{2,1}=0$, \eqref{con2}, and \eqref{con3} are the individual modulation order constraints, where similar arguments apply.

\subsubsection{Power Assignments}\label{sec:2up}
We introduce two \emph{layers} power assignments. The first one is to assign power across different users' modulated symbols at the same time instant within the same sub-block of $\boldsymbol{x}$. The second layer power assignment is performed on top of the first layer power assignment by assigning power across different sub-blocks of $\boldsymbol{x}$. In other words, the first layer power assignment determines the ratio between each user's signal power within the same sub-block while the second layer power assignment determines the power ratio between sub-blocks.

$3a)$ \emph{First layer power assignment}: The power to $v_1[j]$ and $v_2[j]$ for $j=1,\ldots,N_1$ is chosen such that the superimposed symbol satisfies
\begin{align}\label{eq:2upa1}
v_1[j]+\sqrt{2^{m_1}}v_2[j]\in\Lambda_1+\sqrt{2^{m_1}}\Lambda_{2,1},\;j=1,\ldots,N_1,
\end{align}
where the superimposed constellation $\Lambda_1+\sqrt{2^{m_1}}\Lambda_{2,1}$ is a regular QAM with cardinality $2^{m_1+m_{2,1}}$, zero mean, and $d_{\min}(\Lambda_1+\sqrt{2^{m_1}}\Lambda_{2,1})=1$.

$3b)$ \emph{Second layer power assignment}: On top of the first layer power assignment, we assign the power $P_1+P_{2,1}$ and $P_{2,2}$ to the first $N_1$ and the last $N_2-N_1$ symbols, respectively, of $\boldsymbol{x}$ such that the total power constraint \eqref{eq:2userP} is fulfilled. Obviously, $P_{2,2}$ determines the power of the interference-free symbols while $P_1+P_{2,1}$ determines the power of the superimposed symbols. As a result, the transmitted signals for users 1 and 2 after the proposed two layers power assignments are
\begin{align}
x_1[j]=&\eta_1\sqrt{P_1+P_{2,1}}v_1[j],\qquad j=1,\ldots,N_1, \label{eq:x1} \\
x_2[j]=& \left\{ {\begin{array}{*{20}{c}}
\eta_1 \sqrt{2^{m_1}(P_1+P_{2,1})}v_2[j],&j=1,\ldots,N_1\\
\eta_2\sqrt{P_{2,2}}v_2[j], &j=N_1+1,\ldots,N_2\\
\end{array}} \right.,\label{eq:x2}
\end{align}
where $\eta_1 = \sqrt{\frac{6}{2^{m_1+m_{2,1}}-1}}$ and $\eta_2 = \sqrt{\frac{6}{2^{m_{2,2}}-1}}$ are the normalization factors to ensure both $\Lambda_1+\sqrt{2^{m_1}}\Lambda_{2,1}$ and $\Lambda_{2,2}$ have unit energy. The computation of normalization factors follows Section \ref{sec:preliminaries}. For $X_1\overset{\text{unif}}{\sim}\eta_1\sqrt{P_1+P_{2,1}}\Lambda_1$ and $X_{2,1}\overset{\text{unif}}{\sim}\eta_1\sqrt{2^{m_1}(P_1+P_{2,1})}\Lambda_{2,1}$, based on the last equality of \eqref{eq:2userP} and the proposed first layer power assignment introduced in \eqref{eq:2upa1}, we can obtain $P_1$ and $P_{2,1}$ as
\begin{align}
P_1 =&\E_{X_1}[|X_1|^2] \nonumber \\
=& \frac{2^{m_1}-1}{2^{m_1+m_{2,1}} -1}\left(\frac{N_2}{N_1}P - \frac{N_2-N_1}{N_1}P_{2,2}\right), \label{eq:P1}
\end{align}
and
\begin{align}
P_{2,1} =&\E_{X_{2,1}}[|X_{2,1}|^2]\nonumber \\
=& \frac{2^{m_1+m_{2,1}}-2^{m_1}}{2^{m_1+m_{2,1}} -1}\left(\frac{N_2}{N_1}P - \frac{N_2-N_1}{N_1}P_{2,2}\right),\label{eq:P21}
\end{align}
respectively. Thus, when $P_{2,2}$ is given, $P_1$ and $P_{2,1}$ become deterministic. In our scheme, we consider balanced second layer power assignment for each sub-block of superimposed symbol block $\boldsymbol{x}$ such that $P_{2,2}=P_1+P_{2,1}$. This also means that each sub-block of $\boldsymbol{v}_2$ has different power $P_{2,2}\neq P_{2,1}$ as long as $P_1 \neq 0$. This choice allows to strike a balance between the maximum transmission rates among the sub-blocks of $\boldsymbol{x}$. We will show that this choice is good enough for the proposed scheme with QAM and TIN decoding to achieve rate pairs very close to those assume Gaussian and shell codes with perfect SIC and globally optimized $P_1$, $P_{2,1}$ and $P_{2,2}$ for maximizing achievable rate regions.

$3c)$ \emph{Minimum distance}: By looking into the individual constellation while treating the other user's signals as noise, one can see that after the channel effects, i.e., $h_1x_1[j]\in h_1\eta_1\sqrt{P_1+P_{2,1}}\Lambda_1$ and $h_2x_2[j]\in h_2\eta_1\sqrt{P_1+P_{2,1}}\Lambda_{2,1}$, the minimum distance of each constellation satisfies
\begin{align}
d_{\min}&\left(h_1\eta_1\sqrt{P_1+P_{2,1}}\Lambda_1\right)\nonumber \\
=&\sqrt{\frac{6(P_1+P_{2,1})|h_1|^2}{2^{m_1+m_{2,1}}-1}}\overset{\eqref{con1}}{\geq}1, \\
d_{\min}&\left(h_2\eta_1 \sqrt{2^{m_1}(P_1+P_{2,1})}\Lambda_{2,1}\right)\nonumber \\
=&\sqrt{\frac{6(P_1+P_{2,1})|h_2|^2\cdot2^{m_1}}{2^{m_1+m_{2,1}}-1}}\overset{\eqref{con2}}{\geq}1.\label{eq:dmin2}
\end{align}
Notice that for \eqref{eq:dmin2}, the logarithm in \eqref{con2} without 1 inside leads to a constant minimum distance lower bound. Hence, for any $(h_1,h_2)$ satisfying $|h_1|>|h_2|$, both constraints \eqref{con1}, \eqref{con2}, and the proposed power assignments in \eqref{eq:x1}-\eqref{eq:x2} \emph{guarantee constant minimum distance lower bound} for the superimposed constellation and each individual constellation after channel effects and normalization. The constant minimum distance is beneficial to TIN decoding for handling structural interference. It is worth noting that the structural interference comes from the fact that the interfering signal is uniformly distributed over a regular QAM in our design. In contrast, the conventional assumption of using Gaussian input distribution makes the interference Gaussian which is highly unstructured. As for $j=N_1+1,\ldots,N_2$, the constellation $\Lambda_{2,2}$ is already a regular QAM with $d_{\min}(\Lambda_{2,2})=1$ and $x_2[j]$ is interference-free. Thus, the first layer power assignment is not required here. With \eqref{con3}, one can easily verify that $d_{\min}(h_2\eta_2\sqrt{P_{2,2}}\Lambda_{2,2})\geq 1$.

\subsubsection{TIN Decoding}
At the receiver, each user decodes its own messages by treating the other user's signals as noise. Hence, the other user's codebook information is completely unnecessary for the proposed scheme. For user $i$, $i\in\{1,2\}$, the decoder first computes the log-likelihood ratio (LLR) for each bit of the interleaved codeword $\boldsymbol{\tilde{c}}_i$ from the received signals $\boldsymbol{y}_i$ in \eqref{eq:yk}. Then, the LLR sequence is deinterleaved and passed into a soft-input soft-out decoder. The whole decoding process is similar to that in the point-to-point channel. The detailed process can be found in \cite{CIT-019} and is omitted here due to the space limitation.

It is worth noting that even though the proposed scheme can operate well with TIN, with the knowledge of other user's codebook, the possibility of using SIC, when feasible, to achieve a higher rate pair is not precluded. SIC in the traditional sense requires user 1 to receive the whole frame of $\boldsymbol{x}$ with length $N_2\geq N_1$. On the other hand, one may employ early decoding as suggested in \cite{9518265,9838392} such that user 1 can perform SIC once the first $N_1$ symbols of $\boldsymbol{x}$ are received. However, the conditions for enabling early SIC decoding in our case are different from those in \cite{9518265,9838392} which assumes i.i.d. Gaussian signaling. It should be noted that SIC is imperfect under the finite blocklenth constraints and thus can introduce error propagation to the decoding of the desired codeword. Moreover, the delay and computational complexity due to SIC increase with the number of users.

\begin{remark}
From \eqref{eq:P1} and \eqref{eq:P21}, we see that each individual constellation is normalized to satisfy an \emph{average} power constraint. This is commonly adopted in practical communication systems such that the transmit power can be controlled. However, for some input distributions like Gaussian codes and QAM signaling excluding 4-QAM, the average power constraint is not exactly the same as the \emph{maximal} power constraint per codeword introduced in Sections \ref{sec:model} and \ref{sec:2u_const}, e.g., $\sum^{N_1}_{j=1}|x_1[j]|^2 \leq N_1P_1$. As illustrated in Lemma \ref{lem:power_constraint} in Appendix \ref{APP1}, we can reduce the average power by applying an arbitrarily small constant $\delta>0$, e.g., $\E_{X_1}[|X_1|^2]=P_1-\delta$. This ensures that the probability of a sequence of randomly generated coded symbols that violates the maximal power constraint decreases exponentially in its blocklength. This approach was used in various works that assume Gaussian signaling, from early error exponent analysis \cite{1057022} to very recent related works \cite{9518265,9838392}. Following similar steps in \cite[Lemma 1]{9838392}, the total power constraint \eqref{eq:total_power} can also be satisfied. Since $\delta$ is very small and has negligible impacts in our set-up, we still adopt the original notation without $\delta$ for ease of presentation.
\demo
\end{remark}

The above scheme is based on the assumption of $|h_1|>|h_2|$. For the second layer power assignment, the encoding, modulation, and decoding steps remain the same regardless of the channel order. When $|h_2|<|h_1|$, the proposed first layer power assignment simply swaps the arguments between constellations $\Lambda_1$ and $\Lambda_{2,1}$ and their modulation orders in \eqref{eq:2upa1}. That is, we assign the power to $v_1[j]$ and $v_2[j]$ by $v_2[j]+2^{m_{2,1}}v_1[j]$ for $j=1,\ldots,N_1$, such that $d_{\min}(\Lambda_{2,1}+\sqrt{2^{m_{2,1}}}\Lambda_1)=1$, $d_{\min}(h_1\eta_1\sqrt{2^{m_{2,1}}P_1+P_{2,1}}\Lambda_1)\geq 1$, and $d_{\min}(h_2\eta_1\sqrt{P_1+P_{2,1}}\Lambda_{2,1})\geq 1$. Intuitively speaking, the weak user gets assigned with more power while the strong user is assigned with less power in order to ensure their superimposed constellation and the individual constellation after the channel effects have constant minimum distance lower bound.

\subsection{An Extension to $K$-User Case}\label{sec:schemeL}
We generalize the scheme in Section \ref{sec:2user} to the $K$-user case. Since the encoding and decoding are  single-user based as in the two-user case, we only focus on the modulation mapping and power assignments. We emphasize again that each user only uses a single channel code. We reserve $k\in\{1,\ldots,K\}$ to be the user index or the sub-block index of the superimposed symbol block $\boldsymbol{x}$, and $k'\in\{1,\ldots,k\}$ to be the sub-block index of user $k$'s modulated symbol block $\boldsymbol{v}_k$ or the symbol block after power assignment $\boldsymbol{x}_k$. The blocklengths of $K$ users follow the assumption in Section \ref{sec:model}, which is $N_1\leq\ldots\leq N_K$ without loss of generality. From \eqref{eq:yk}, we note that the superimposed symbol block $\boldsymbol{x}$ can be divided into $K$ sub-blocks of lengths $N_1,N_2-N_1,\ldots,N_K-N_{K-1}$, respectively. Moreover, the $k$-th sub-block of $\boldsymbol{x}$ has length $N_k-N_{k-1}$ with $N_0=0$ and is the superposition of the modulated symbols of users $k,k+1,\ldots,K$. For these $K-k+1$ users, we assume that their channel gains follow some order $|h_{g_1}|>|h_{g_2}|>\ldots>|h_{g_{K-k+1}}|$, where $h_{g_1}$ represents the channel that has the largest gain among $h_k,\ldots,h_K$, $h_{g_2}$ represents the channel with the second largest channel gain, etc. One can think of $[g_1,\ldots,g_{K-k+1}]$ as a permutation of user index vector $[k,\ldots,K]$.

\subsubsection{Modulation Mapping}\label{sec:kmp}
For user $k$, its transmitted packet experiences other $K-1$ users' heterogeneous interference as shown in \eqref{eq:yk}. To handle heterogeneous interference, user $k$ uses $k$ sets of constellations $\Lambda_{k,1},\ldots,\Lambda_{k,k}$. Moreover, user $k$'s modulated symbol block $\boldsymbol{v}_k$ is divided into $k$ sub-blocks, where the $k'$-th sub-block is associated with $\Lambda_{k,k'}$ for $k'\in\{1,\ldots,k\}$. $\Lambda_{k,k'}$ is a regular Gray labeled QAM with zero mean, minimum distance 1 and cardinality $2^{m_{k,k'}}=2^{\log|\Lambda_{k,k'}|}$. We denote by $\boldsymbol{\tilde{c}}_k$ the interleaved codeword of length $n_k$ for user $k$. The interleaved codeword is divided into $k$ sub-sequences, i.e., $\tilde{\boldsymbol{c}}_k=[\tilde{\boldsymbol{c}}_{k,1},\ldots,\tilde{\boldsymbol{c}}_{k,k}]$, where $\tilde{\boldsymbol{c}}_{k,k'}$ has $(N_{k'}-N_{k'-1})m_{k,k'}$ bits for $k'\in\{1,\ldots,k\}$. Then, $\tilde{\boldsymbol{c}}_k$ is mapped to the symbol sequence $\boldsymbol{v}_{k}=[\boldsymbol{v}_{k,1},\ldots,\boldsymbol{v}_{k,k}]\in\mathbb{C}^{N_{k}}$, in which $\tilde{\boldsymbol{c}}_{k,k'}$ is mapped to $\boldsymbol{v}_{k,k'}=\left[v_k[N_{k'-1}+1],\ldots,v_k[N_{k'}]\right]\in\mathbb{C}^{N_{k'}-N_{k'-1}}$ with $v_k[j] \in \Lambda_{k,k'}$ for $j=N_{k'-1}+1,\ldots,N_{k'}$. As a result, the relationship between codeword length $n_k$ and modulated symbol length $N_k$ for user $k$ satisfies
\begin{align}\label{eq:lengthnk}
n_k = \sum\nolimits^k_{k'=1}(N_{k'}-N_{k'-1})m_{k,k'}.
\end{align}
We provide an example to illustrate the above mapping.

\begin{example}\label{example1}
Consider $K=3$, $N_1=200$, $N_2=1000$, and $N_3=2000$. For user 3, we assume that $\Lambda_{3,1}$, $\Lambda_{3,2}$, and $\Lambda_{3,3}$ are 4-QAM, 16-QAM, and 4-QAM, respectively. The codeword length of user 3 is 5600. For the first 400 bits of the codeword $\boldsymbol{\tilde{c}}_3$, every 2 bits are mapped to a constellation point of $\Lambda_{3,1}$, leading to 200 4-QAM symbols. For the next 3200 bits of $\boldsymbol{\tilde{c}}_3$, every 4 bits are mapped to a constellation point of $\Lambda_{3,2}$, resulting in 800 16-QAM symbols. Finally, every 2 bits of the last 2000 bits of $\boldsymbol{\tilde{c}}_3$ are mapped to a constellation point of $\Lambda_{3,3}$ and become 1000 4-QAM symbols.
\demo
\end{example}

We then introduce the constraints on the modulation orders based on sub-block power constraints and channels. We denote by $\boldsymbol{x}_k$ the symbol block which is after applying power assignments on $\boldsymbol{v}_k$ for user $k\in\{1,\ldots,K\}$. The symbol block $\boldsymbol{x}_k$ can also be decomposed into $k$ sub-blocks the same way as for $\boldsymbol{v}_k$. The $k'$-th sub-block of $\boldsymbol{x}_k$ for $k'\in\{1,\ldots,k\}$ satisfies the following power constraint
\begin{align}\label{eq:sub_powerk}
\frac{1}{N_{k'}-N_{k'-1}}\sum\nolimits_{j=N_{k'-1}+1}^{N_k'}|x_k[j]|^2 \leq& P_{k,k'}.
\end{align}
The relationship between the sub-block power constraint \eqref{eq:sub_powerk} and the individual power constraint  \eqref{eq:ind_power} for user $k$ satisfies
\begin{align}
\frac{1}{N_k}\sum\nolimits_{j=1}^{N_k}|x_k[j]|^2 \leq \sum\nolimits^k_{k'=1}\frac{N_{k'}-N_{k'-1}}{N_k}P_{k,k'} = P_k.
\end{align}
Furthermore, the relationship between each sub-block power constraint and the total power constraint \eqref{eq:total_power} satisfies
\begin{align}\label{eq:kuserP}
\sum\nolimits^K_{k=1}\frac{N_{k}-N_{k-1}}{N_K}\left(\sum\nolimits^K_{i=k}P_{i,k}\right)= P.
\end{align}
From \eqref{eq:sub_powerk}, we obtain the power constraint for the $k$-th sub-block of the superimposed symbol block $\boldsymbol{x}$ for $k\in\{1,\ldots,K\}$ as
\begin{align}\label{eq:sup_pc1}
\frac{1}{N_{k}-N_{k-1}}&\sum\nolimits^{N_{k}}_{j=N_{k-1}+1}(|x'_1[j] |^2+\ldots+|x'_K[j] |^2) \nonumber \\
\leq& \sum\nolimits^K_{l=k}P_{l,k}.
\end{align}
To construct the $k$-th sub-block of $\boldsymbol{x}$, $k\in\{1,\ldots,K\}$, we propose the following criteria on the modulation associated with each sub-block of $\boldsymbol{v}_k$ based on the channel order introduced at the beginning of Section \ref{sec:schemeL} and the sub-block power constraint \eqref{eq:sup_pc1}
\begin{align}\label{conk}
\sum\nolimits^{K-k+1}_{i'=i}m_{g_{i'},k} \leq& \left\lfloor\log \left(6\sum\nolimits^{K}_{l=k}P_{l,k}|h_{g_i}|^2\right)\right\rfloor, \nonumber\\
&i=1,\ldots,K-k+1,
\end{align}
where $h_{g_i}$ represents the channel among $h_k,h_{k+1},\ldots,h_K$ that is ranked in the $i$-th position based on the channel gain and $m_{g_i,k}$ is modulation order associated with the user that has channel $h_{g_i}$. The RHS of \eqref{conk} is reminiscent of the single-user capacity of the user with the largest signal-to-noise-ratio (SNR) among users $k,k+1,\ldots,K$ with the first $K-k+2-i$ smallest channel gains. It also introduces the individual modulation order constraint with $m_{g_{i},k} \leq \lfloor\log (\sum^{K}_{l=k}P_{l,k}|h_{g_i}|^2)\rfloor$. As we have seen in the two-user case, this constraint is closely related to the minimum distance of the superimposed and individual constellations, which will be explained in Section \ref{sec:paK}. We emphasize that the modulation orders should be chosen such that their sum is close to the RHS of \eqref{conk} in order to obtain higher rates.

\subsubsection{Power Assignments}\label{sec:paK}
Similar to the two-user case, we use two layers power assignments. From Section \ref{sec:kmp}, we know that the symbols in $k'$-th sub-block of $\boldsymbol{v}_k$ are drawn from $\Lambda_{k,k'}$ such that $v_k[j] \in \Lambda_{k,k'}$ for $j=N_{k'-1}+1,\ldots,N_{k'}$ and $k'=1,\ldots,k$. The modulated symbols of users $k',k'+1,\ldots,K$ are superimposed to form the $k'$-th sub-block of $\boldsymbol{x}$. Assume that user $k$'s channel is ranked the $i$-th position for the channel gain for some $g_i\in \{g_1,\ldots,g_{K-k'+1}\}$, i.e., $h_{k}=h_{g_i}$, among $h_{k'},h_{k'+1},\ldots,h_K$.

\emph{2a) First layer power assignment}: We assign the power to the modulated symbol for user $k$ in the $k'$-th sub-block of $\boldsymbol{v}_{k}$ such that
\begin{align}\label{eq:fpk}
\sqrt{2^{\sum^{i-1}_{i'=1}m_{g_{i'},k'}}}v_{k}[j]\in&\sqrt{2^{\sum^{i-1}_{i'=1}m_{g_{i'},k'}}}\Lambda_{k,k'}, \nonumber\\ &j=N_{k'-1}+1,\ldots,N_{k'},
\end{align}
where $m_{g_{i'},k'}=\log |\Lambda_{g_{i'},k'}|$ and $\Lambda_{g_{i'},k'}$ is constellation used in the $k'$-th sub-block of $\boldsymbol{x}$, whose user has a smaller channel gain than user $k$. As a result, the superimposed constellation associated with the $k'$-th sub-block of $\boldsymbol{x}$, i.e., $\Lambda_{g_1,k'}+\sum^{K-k'+1}_{i=2}\sqrt{2^{\sum^{i-1}_{i'=1}m_{g_{i'},k'}}}\Lambda_{g_{i},k'}$, is a regular QAM with zero mean and minimum distance 1 and cardinality $2^{\sum^{K-k'+1}_{i'=1}m_{g_{i'},k'}}$. We use the following example to illustrate the main idea behind the above proposed scheme.
\begin{example}\label{example2}
Consider $K=3$, $N_1 \leq N_2 \leq N_3$, and $|h_2|>|h_1|>|h_3|$. After the first layer power assignment, the sub-blocks of $\boldsymbol{v}_1$, $\boldsymbol{v}_2$, and $\boldsymbol{v}_3$ become $(\sqrt{2^{m_{2,1}}}v_1[j],v_2[j],\sqrt{2^{m_{2,1}+m_{1,1}}}v_3[j])$ for $j=1,\ldots,N_1$, $(0,v_2[j],\sqrt{2^{m_{2,2}}}v_3[j])$ for $j=N_1+1,\ldots,N_2$, and $(0,0,v_3[j])$ for $j=N_2+1,\ldots,N_3$, where 0 simply means no symbols.
\demo
\end{example}

\emph{2b) Second layer power assignment}: Further to \eqref{eq:fpk}, the $k'$-th sub-block of $\boldsymbol{x}_k$ after the second layer power assignment for $k'\in\{1,\ldots,k\}$ is
\begin{align}\label{eq:xk}
x_k[j]=&\eta_{k'}\sqrt{2^{\sum^{i-1}_{i'=1}m_{g_{i'},k'}}\left(\sum\nolimits^K_{l=k'}P_{l,k'}\right)}v_k[j], \nonumber \\ &j=N_{k'-1}+1,\ldots,N_{k'},
\end{align}
where $\sum^K_{l=k'}P_{l,k'}$ is the power constraint on the $k'$-th sub-block of $\boldsymbol{x}$ following \eqref{eq:sup_pc1} and $\eta_{k'}=\sqrt{\frac{6}{2^{\sum^{K-k'+1}_{i'=1}m_{g_{i'},k'}}-1}}=\sqrt{\frac{6}{2^{\sum^{K}_{l=k'}m_{k',k'}}-1}}$ is the normalization factor to ensure that the superimposed constellation $\Lambda_{g_1,k'}+\sum^{K-k'+1}_{i=2}\sqrt{2^{\sum^{i-1}_{i'=1}m_{g_{i'},k'}}}\Lambda_{g_{i},k'}$ has unit energy. Given the ratio between the power constraint on each sub-block of $\boldsymbol{x}$, $\sum^K_{i=1}P_{i,1},\sum^K_{i=2}P_{i,2},\ldots,P_{K,K}$ together with the first layer power assignment, the power of all sub-blocks, $P_{1,1},P_{2,1},P_{2,2},\ldots,P_{K,K}$ become deterministic. Similar to the two-user case, we consider equal power among the sub-block of superimposed coded symbols, i.e., $\sum^K_{i=1}P_{i,1}=\sum^K_{i=2}P_{i,2}=\ldots=P_{K,K}$. Thus, the superimposed symbol block $\boldsymbol{x}$ has the same power across its symbols while each sub-block of user $k$'s symbol block $\boldsymbol{x}_k$ has a different power. As a result, the $k'$-sub-block power of user $k$ is obtained as
\begin{align}
P_{k,k'}=\E_{X_{k,k'}}[|X_{k,k'}|^2] = \frac{2^{\sum^{i-1}_{i'=1}m_{g_{i'},k'}}(2^{m_{g_i},k'}-1)}{2^{\sum^{K-k'+1}_{i'=1}m_{g_{i'},k'}}-1}P.
\end{align}
It is interesting to see that the proposed power assignments are not based on individual blocklength $N_k$. This means that under the same channel condition and as long as the blocklength order does not change, our power assignments do not need to be changed.

\emph{2c) Minimum distance}: By looking at the individual constellation after normalization and channel effects while treating other users' signals as noise, it can be verified that the minimum distance satisfies
\begin{align}
&d_{\min}\left(h_{k}\eta_{k'}\sqrt{2^{\sum^{i-1}_{i'=1}m_{g_{i'},k'}}\sum\nolimits^K_{l=k'}P_{l,k'}}\Lambda_{k,k'}\right)\nonumber \\
=&\sqrt{\frac{6\cdot2^{\sum^{i-1}_{i'=1}m_{g_{i'},k'}}|h_{g_i}|^2\sum^K_{l=k'}P_{l,k'}}{2^{\sum^{K-k'+1}_{i'=1}m_{g_{i'},k'}}-1}}\overset{\eqref{conk}}{\geq}1,
\end{align}
Hence, given any channel order, the proposed power assignments ensure that the superimposed constellation for each sub-block of $\boldsymbol{x}$ and the individual constellation for each sub-block of $\boldsymbol{v}_k$ after normalization and channel effects have constant minimum distance lower bound.

\section{Finite Blocklength Achievable Rate Analysis}\label{sec:FBL}
In this section, we derive the second-order achievable rate of the downlink BC with discrete signaling and TIN with given blocklength and error probability constraints. The derived achievable rate will be used to guide the design of channel code parameters and also estimate the block error probability. As in Section \ref{sec:model}, we assume $N_1\leq \ldots\leq N_K$ without loss of generality. The channel order is not required here due to the fact that TIN decoding is adopted.

\subsection{Two-User Case}\label{sec:analysis2}
For ease of understanding, we first present the analysis of the two-user downlink BC. We define the normalized constellations after power assignments for $\boldsymbol{x}_1$ and two sub-blocks of $\boldsymbol{x}_2$ as $\mathcal{X}_1$, $\mathcal{X}_{2,1}$, and $\mathcal{X}_{2,2}$, respectively.

\subsubsection{Achievable Rate of User 1}
We first analyze the information density, which is the key to our second-order achievable rate approximation. Based on the definition of information density in \cite{5452208}, the information density of user 1 is derived in \eqref{eq:u1id_basic}-\eqref{eq:u1id_basic1} at the top of page 9,
\begin{figure*}[t]
\begin{align}
i(X_1^{[N_1]};Y_1^{[N_1]}) =& \sum\nolimits_{j=1}^{N_1}i(X_1[j];Y_1[j])
=\sum\nolimits_{j=1}^{N_1}\log \left(\frac{P_{X_1|Y_1}(x_1[j]|y_1[j])}{P_{X_1}(x_1[j])}\right)\label{eq:u1id_basic}
\\
=&\sum\nolimits_{j=1}^{N_1} \log \left(\frac{\sum\limits_{x_2[j] \in \mathcal{X}_{2,1}}P_{Y_1|X_1,X_2}(y_1[j]|x_1[j],x_2[j])}{\sum\limits_{x_1[j] \in
\mathcal{X}_1}\sum\limits_{x_2[j] \in \mathcal{X}_{2,1}}P_{Y_1|X_1,X_2}(y_1[j]|x_1[j],x_2[j])P_{X_1}(x_1[j])}\right), \label{eq:u1id_basic1}
\end{align}
\hrule
\end{figure*}
where we note that $X_1[j], X_2[j]$, and $Y_1[j]$ are i.i.d. for $j=1,\ldots,N_1$ and thus $i(X_1[j];Y_1[j])$ is also i.i.d., $P_{Y_1|X_1,X_2}(y_1[j]|x_1[j],x_{2}[j])=\frac{1}{\pi}e^{-|y_1[j]-h_1(x_1[j]+x_{2}[j])|^2}$, and $P_{X_1}(x_1[j]) = \frac{1}{|\mathcal{X}_1|}$ and $P_{X_2}(x_2[j]) = \frac{1}{|\mathcal{X}_{2,1}|}$ due to uniform input distributions of $x_1[j]$ and $x_2[j]$, respectively, for $j=1,\ldots,N_1$. With the information density, we derive the mutual information for user 1 as
\begin{align}
I(X_1^{[N_1]};Y_1^{[N_1]}) \overset{\eqref{eq:u1id_basic}}=&\displaystyle \sum\nolimits_{j=1}^{N_1}\E\left[i(X_1[j];Y_1[j])\right] \nonumber \\
=&N_1\displaystyle \E[i(X_1;Y_1)]=N_1I(X_1;Y_1) \label{eq:NI1},
\end{align}
where the two last equalities in \eqref{eq:NI1} follows because $i(X_1[j];Y_1[j])$ is i.i.d. and thus the symbol index $[j]$ in the expectation can be dropped. The mutual information $I(X_1^{[N_1]};Y_1^{[N_1]})$ reminds the reader that it is under TIN as opposed to $I(X_1^{[N_1]};Y_1^{[N_1]}|X_2^{[N_1]})$ with SIC. Further to \eqref{eq:NI1}, we derive $I(X_1;Y_1)$ in \eqref{eq:u1I} at the top of page 9.
\begin{figure*}[t]
\begin{align}\label{eq:u1I}
I(X_1;Y_1)= \log|\mathcal{X}_1|
-\frac{1}{|\mathcal{X}_{1}|\cdot|\mathcal{X}_{2,1}|}\sum_{x_{1}\in\mathcal{X}_{1}}\sum_{x_{2,1}\in\mathcal{X}_{2,1}}\E_{Z_1}\left[\log \left(\frac{\sum\limits_{x'_1\in \mathcal{X}_1}\sum\limits_{x'_{2,1}\in \mathcal{X}_{2,1}}e^{-|Z_1+h_1(x_1-x'_1+x_{2,1}-x'_{2,1})|^2}}{\sum\limits_{x'_{2,1}\in \mathcal{X}_{2,1}}e^{-|Z_1+h_1(x_{2,1}-x'_{2,1})|^2}}\right) \right].
\end{align}
\hrule
\end{figure*}

Next, we derive the dispersion function $V(X_1^{[N_1]};Y_1^{[N_1]})$ which will be used in the derivation of second-order achievable rate for user 1
\begin{align}
V(X_1^{[N_1]};Y_1^{[N_1]})
\overset{\eqref{eq:u1id_basic}}{=}&\displaystyle\mathop{\text{Var}}\left[\sum\nolimits_{j=1}^{N_1}i(X_1[j];Y_1[j])\right]\nonumber\\
=&\sum\nolimits_{j=1}^{N_1}\displaystyle\mathop{\text{Var}}[i(X_1[j];Y_1[j])] \label{eq:u1Vint}\\
=& N_1V(X_1;Y_1),\label{eq:N1V1}
\end{align}
where the last equality in \eqref{eq:u1Vint} holds because $x_1[j]$ and $x_1[j']$ are independent and $y_1[j]$ and $y_1[j']$ are independent for any $j\neq j'$ and $j,j'\in \{1,\ldots,N_1\}$ and \eqref{eq:N1V1} follows that $X_1[j]$ and $Y_1[j]$ are i.i.d. for $j=1,\ldots,N_1$ and thus the symbol index $[j]$ can be dropped. Further to \eqref{eq:N1V1}, we have $V(X_1;Y_1)$ shown in \eqref{eq:u1V} at the top of page 9.
\begin{figure*}[t]
\begin{align}
V(X_1;Y_1)= &\mathop{\E}[(i(X_1;Y_1))^2]-\left( \mathop{\E}[i(X_1;Y_1)]\right)^2 \\
=&\frac{1}{|\mathcal{X}_{1}|\cdot|\mathcal{X}_{2,1}|}\sum_{x_{1}\in\mathcal{X}_{1}}\sum_{x_{2,1}\in\mathcal{X}_{2,1}}\E_{Z_1}\left[\left(\log\left( \frac{\sum\limits_{x'_1\in \mathcal{X}_1}\sum\limits_{x'_{2,1}\in \mathcal{X}_{2,1}}e^{-|Z_1+h_1(x_1-x'_1+x_{2,1}-x'_{2,1})|^2}}{\sum\limits_{x'_{2,1}\in \mathcal{X}_{2,1}}e^{-|Z_1+h_1(x_{2,1}-x'_{2,1})|^2}}\right)\right)^2 \right]\nonumber\\
&-\left(\frac{1}{|\mathcal{X}_{1}|\cdot|\mathcal{X}_{2,1}|}\sum_{x_{1}\in\mathcal{X}_{1}}\sum_{x_{2,1}\in\mathcal{X}_{2,1}}\E_{Z_1}\left[\log \left(\frac{\sum\limits_{x'_1\in \mathcal{X}_1}\sum\limits_{x'_{2,1}\in \mathcal{X}_{2,1}}e^{-|Z_1+h_1(x_1-x'_1+x_{2,1}-x'_{2,1})|^2}}{\sum\limits_{x'_{2,1}\in \mathcal{X}_{2,1}}e^{-|Z_1+h_1(x_{2,1}-x'_{2,1})|^2}}\right) \right]\right)^2 \label{eq:u1V}.
\end{align}
\hrule
\end{figure*}

We then have the following proposition for the second-order achievable rate of user 1.
\begin{proposition}\label{prof:u1}
Define $\epsilon_1$ to be the upper bound on the average TIN decoding error probability of user 1. For the channel model in \eqref{eq:yk}, user 1's achievable rate by treating user 2's signals as noise is bounded by
\begin{align}\label{eq:u1rate}
R_1 \leq& I(X_1;Y_1)-\sqrt{\frac{V(X_1;Y_1)}{N_1}}Q^{-1}\left(\epsilon_1\right)+O\left(\frac{\log N_1}{N_1}\right),
\end{align}
where $I(X_1;Y_1)$ is given in \eqref{eq:u1I} and $V(X_1;Y_1)$ is given in \eqref{eq:u1V}.
\end{proposition}
\begin{IEEEproof}
Please see Appendix \ref{app:u1}.
\end{IEEEproof}

Since user 1 has the shortest symbol blocks, each intended symbol for user 1 experiences the same interference statistics, which is similar to the homogeneous blocklength case. However, the interference experienced by user 2 behaves differently from user 1 as we will see in the next section.

\subsubsection{Achievable Rate of User 2}\label{sec:u2}
Since $\boldsymbol{x}_2$ will be partially interfered, $X_2[j]$ and $Y_2[j]$ are i.i.d. when either $j = 1,\ldots,N_1$ or $j=N_1+1,\ldots,N_2$ whereas $X_2[j]$ $(Y_2[j])$ and $X_2[j']$ $(Y_2[j'])$ are \emph{not necessarily} identically distributed for $j\in\{1,\ldots,N_1\}$ and $j'\in\{N_1+1,\ldots,N_2\}$. In this case, we let $X_{2,1}$ ($Y_{2,1}$) represent the random variables $X_2[j]$ ($Y_2[j]$) for $j \in \{1, . . . ,N_1\}$ and let $X_{2,2}$ ($Y_{2,2}$) represent the random variables $X_2[j]$ ($Y_2[j]$) for $j \in \{N_1 + 1, . . . ,N_2\}$. The information density for user 2 is derived in \eqref{eq:u2id_basic}-\eqref{eq:u2id} at the top of page 10,
\begin{figure*}
\begin{align}
i(X_2^{[N_2]};Y_2^{[N_2]}) =&\sum_{j=1}^{N_1}\log \left(\frac{P_{X_{2,1}|Y_{2,1}}(x_2[j]|y_2[j])}{P_{X_{2,1}}(x_2[j])}\right)+\sum_{j=N_1+1}^{N_2}\log \left( \frac{P_{X_{2,2}|Y_{2,2}}(x_2[j]|y_2[j])}{P_{X_{2,2}}(x_2[j])}\right) \label{eq:u2id_basic} \\
=&\sum_{j=1}^{N_1}\log \left( \frac{\sum_{x_1[j]\in\mathcal{X}_1}P_{Y_{2,1}|X_1,X_{2,1}}(y_2[j]|x_2[j],x_1[j])}{\sum_{x_2[j]\in\mathcal{X}_{2,1}}\sum_{x_1[j]\in\mathcal{X}_1}P_{Y_{2,1}|X_1,X_{2,1}}(y_2[j]|x_2[j],x_1[j])P_{X_{2,1}}(x_2[j])}\right)\nonumber \\
&+\sum_{j=N_1+1}^{N_2}\log\left( \frac{P_{Y_{2,2}|X_{2,2}}(y_2[j]|x_2[j])}{\sum_{x_2[j]\in\mathcal{X}_{2,2}}P_{Y_{2,2}|X_{2,2}}(y_2[j]|x_2[j])P_{X_{2,2}}(x_2[j])} \right), \label{eq:u2id}
\end{align}
\hrule
\end{figure*}
where $P_{Y_{2,1}|X_1,X_{2,1}}(y_2[j]|x_2[j],x_1[j])=\frac{1}{\pi}e^{-|y_2[j]-h_2(x_1[j]+x_{2}[j])|^2}$ and $P_{X_{2,1}}(x_2[j]) = \frac{1}{|\mathcal{X}_{2,1}|}$ for $j=1,\ldots,N_1$ and $P_{Y_{2,2}|X_{2,2}}(y_2[j]|x_2[j])=\frac{1}{\pi}e^{-|y_2[j]-h_2x_{2}[j])|^2}$ and $P_{X_{2,2}}(x_2[j]) = \frac{1}{|\mathcal{X}_{2,2}|}$ for $j=N_1+1,\ldots,N_2$. With \eqref{eq:u2id}, we obtain the mutual information under TIN for user 2
\begin{align}\label{eq:NI2}
I(X_2^{[N_2]};Y_2^{[N_2]})=&N_1I(X_{2,1};Y_{2,1}) \nonumber \\
&+(N_2-N_1)I(X_{2,2};Y_{2,2}),
\end{align}
where $I(X_{2,1};Y_{2,1})$ can be obtained from \eqref{eq:u1I} by swapping the arguments between users 1 and 2 while $I(X_{2,2};Y_{2,2})$ is the mutual information of the single user channel. Next, we derive the dispersion function as
\begin{align}\label{eq:N2V2}
V(X_2^{[N_2]};Y_2^{[N_2]}) = &N_1V(X_{2,1};Y_{2,1})\nonumber\\
&+(N_2-N_1)V(X_{2,2};Y_{2,2}),
\end{align}
where $V(X_{2,1};Y_{2,1})$ can be obtained from \eqref{eq:u1V} by swapping the arguments between users 1 and 2 while $V(X_{2,2};Y_{2,2})$ is the dispersion of the single user channel.

Having derived the mutual information and dispersion, we have the following proposition for the second-order achievable rate of user 2.
\begin{proposition}\label{prof:u2}
Define $\epsilon_2$ to be the upper bound on the average TIN decoding error probability of user 2. For the channel model in \eqref{eq:yk}, user 2's achievable rate by treating user 1's signals as noise is bounded by
\begin{align}\label{eq:u2rate}
&R_2 \leq \frac{N_1}{N_2}I(X_{2,1};Y_{2,1})+\frac{N_2-N_1}{N_2}I(X_{2,2};Y_{2,2}) \nonumber\\
&-\frac{\sqrt{N_1V(X_{2,1};Y_{2,1})+(N_2-N_1)V(X_{2,2};Y_{2,2})}}{N_2}Q^{-1}\left(\epsilon_2\right)\nonumber\\
&+O\left(\frac{\log N_2}{N_2}\right).
\end{align}
\end{proposition}
\begin{IEEEproof}
Please see Appendix \ref{app:u2}.
\end{IEEEproof}

The impacts of the length of interfering symbols on user 2's achievable rate are clearly shown in \eqref{eq:u2rate}. This is different from the homogeneous blocklength case for which a single signal-to-interference-plus-noise ratio (SINR), i.e., $\frac{P_2|h_2|^2}{P_1|h_2|^2+1}$, could not capture the effects of partially interfered symbols.

\subsubsection{Modulation and Code Design}
With the derived achievable rates, we can design the modulations and channel codes for the proposed schemes. First, consider the blocklength $(N_1,N_2)$ and error probability $(\epsilon_1,\epsilon_2)$ requirements for both users. We design the modulations $(\Lambda_1,\Lambda_{2,1},\Lambda_{2,2})$ whose orders satisfying \eqref{con1} such that the achievable rate pair computed by using \eqref{eq:u1rate} and \eqref{eq:u2rate} reach a target rate pair $(R_1,R_2)$. Note that given the modulations, the power assignments become deterministic according to Section \ref{sec:2up}. Moreover, users 1 and 2's codeword lengths satisfy $(n_1,n_2)=(N_1m_1,N_1m_{2,1}+(N_2-N_1)m_{2,2})$ according to \eqref{u1n1} and \eqref{u2n2} in Section \ref{sec:2u_const}. To match users 1 and 2's transmission rates with their corresponding achievable rates, i.e., $(R_1,R_2) =  (\frac{k_1}{n_1}m_1,\frac{k_2}{n_2}(\frac{N_1}{N_2}m_{2,1}+\frac{N_2-N_1}{N_2}m_{2,2}))$, the information lengths of users 1 and 2's channel codes are obtained as $(k_1,k_2)=(R_1N_1,R_2N_2)$. The problem can now be converted into designing good point-to-point codes with the specified information and codeword lengths.

\subsection{$K$-User Case}
We generalize the second-order achievable rate analysis to the $K$-user case. From Section \ref{sec:schemeL}, we note that for user $k$, $k\in \{1,\ldots,K\}$, its symbol block $\boldsymbol{x}_k$ can be decomposed into $k$ sub-blocks. The superimposed symbol block $x[j]$ and received symbol block $y_k[j]$ for $j=1,\ldots,N_k$ can also be decomposed into $k$ sub-blocks the same ways as for $\boldsymbol{x}_k$. It is easy to see that the channel input $X_k[j]$ and output $Y_k[j]$ are i.i.d. for $j=N_{k-1}+1,\ldots,N_k$, where we set $N_0=0$. Hence, we drop the index $[j]$ and use $X_{k,j}$ and $Y_{k,j}$ to represent the channel input and output for the $j$-th sub-block and $j\in\{1,\ldots,k\}$. Moreover, $X_{k,j}$ is uniformly distributed over a constellation $\mathcal{X}_{k,j}$. By following the steps in Section \ref{sec:analysis2}, we derive the second-order achievable rate for user $k$ under TIN decoding and state the result in the following proposition.

\begin{proposition}\label{prop:uk}
Define $\epsilon_k$ to be upper bound on the average TIN decoding error probability of user $k$ for $k\in\{1,\ldots,K\}$. For the channel model in \eqref{eq:yk}, user $k$'s achievable rate by treating users $1,\ldots,k-1,k,\ldots,K$'s signals as noise is bounded by
\begin{align}\label{eq:uk_rate}
R_k \leq& \frac{\sum^k_{j=1}(N_j-N_{j-1})I(X_{k,j};Y_{k,j})}{N_k}\nonumber \\
&-\frac{\sqrt{\sum^k_{j=1}(N_j-N_{j-1})V(X_{k,j};Y_{k,j})}}{N_k}Q^{-1}\left(\epsilon_k\right) \nonumber\\
&+O\left(\frac{\log N_k}{N_k}\right),
\end{align}
where $I(X_{k,j};Y_{k,j})$ and $V(X_{k,j};Y_{k,j})$ are shown in \eqref{eq:Ik} and \eqref{eq:Vk}, respectively, at the top of page 11.
\begin{figure*}
\begin{align}
I(X_{k,j};Y_{k,j}) =& \log|\mathcal{X}_{k,j}|
-\frac{1}{\prod\limits^K_{i=j} |\mathcal{X}_{i,j}|}\sum_{x_{j,j},\ldots,x_{K,j}}\E_{Z_k}\left[\log\left( \frac{\sum\limits_{x'_{j,j},\ldots,x'_{K,j}}e^{-\left|Z_k+h_k\sum^K_{i=j}(x_{i,j}-x'_{i,j})\right|^2}}{\sum\limits_{x'_{j,j},\ldots,x'_{k-1,j},x'_{k+1,j},\ldots,x'_{K,j}}e^{-\left|Z_k+h_k\sum^K_{i =j,i\neq k}(x_{i,j}-x'_{i,j})\right|^2}}\right) \right],\label{eq:Ik}\\
V(X_{k,j};Y_{k,j})
=&\frac{1}{\prod\limits^K_{i=j} |\mathcal{X}_{i,j}|}\sum_{x_{j,j},\ldots,x_{K,j}}\E_{Z_k}\left[\left(\log\left( \frac{\sum\limits_{x'_{j,j},\ldots,x'_{K,j}}e^{-\left|Z_k+h_k\sum^K_{i=j}(x_{i,j}-x'_{i,j})\right|^2}}{\sum\limits_{x'_{j,j},\ldots,x'_{k-1,j},x'_{k+1,j},\ldots,x'_{K,j}}e^{-\left|Z_k+h_k\sum^K_{i =j,i\neq k}(x_{i,j}-x'_{i,j})\right|^2}}\right)\right)^2 \right] \nonumber\\
&-\left(\frac{1}{\prod\limits^K_{i=j} |\mathcal{X}_{i,j}|}\sum_{x_{j,j},\ldots,x_{K,j}}\E_{Z_k}\left[\log\left( \frac{\sum\limits_{x'_{j,j},\ldots,x'_{K,j}}e^{-\left|Z_k+h_k\sum^K_{i=j}(x_{i,j}-x'_{i,j})\right|^2}}{\sum\limits_{x'_{j,j},\ldots,x'_{k-1,j},x'_{k+1,j},\ldots,x'_{K,j}}e^{-\left|Z_k+h_k\sum^K_{i =j,i\neq k}(x_{i,j}-x'_{i,j})\right|^2}}\right) \right]\right)^2,\label{eq:Vk}
\end{align}
\hrule
\end{figure*}
\end{proposition}
\begin{IEEEproof}
Please see Appendix \ref{app:uk}.
\end{IEEEproof}

With the derived achievable rate $R_k$ under symbol length constraint $N_k$, we can find the parameters of user $k$'s channel code by using \eqref{eq:lengthnk} in Section \ref{sec:schemeL} such that the information $k_k$ and codeword length $n_k$ satisfy $R_k=\frac{k_k}{N_k}=\frac{k_k}{n_k}\frac{\sum^k_{k'=1}(N_{k'}-N_{k'-1})m_{k,k'}}{N_k}$. For \eqref{eq:Ik} and \eqref{eq:Vk}, we can further define the superimposed symbol and constellation, i.e., $x_{k,j,\Sigma}\triangleq\sum^K_{i=j}x_{i,j}\in \mathcal{X}_{k,j,\Sigma}\triangleq\bigcup^K_{i=j}\mathcal{X}_{i,j}$. It is worthwhile to note that the numerator inside $\E_{Z_k}[\log(.)]$ is closely related to $d_{\min}(\mathcal{X}_{k,j,\Sigma})$ while the denominator is related to $d_{\min}(\mathcal{X}_{k,j,\Sigma} \setminus \mathcal{X}_{k,j})$. Since $d_{\min}(\mathcal{X}_{k,j,\Sigma} \setminus \mathcal{X}_{k,j}) \geq d_{\min}(\mathcal{X}_{k,j,\Sigma})$, it is important to ensure that the minimum distance of the superimposed constellation will not vanish. Hence, in Section \ref{sec:scheme}, we design our scheme to guarantee a constant minimum distance lower bound for the superimposed constellation of each sub-block of $\boldsymbol{x}$.

\section{Simulation Results}\label{sec:sim}

\subsection{Achievable Rate}

We present some design examples to show the performance of the proposed scheme with QAM and TIN. We define $\SNR_k \triangleq P|h_k|^2$ for user $k\in\{1,\ldots,K\}$. We first consider a two-user downlink BC, where $(\SNR_1,\SNR_2)=(24,12)$ in dB, $N_1=N_2=N=200$, and $\epsilon_1=\epsilon_2=10^{-6}$. Under equal blocklengths, we have $\Lambda_{2,1}=\Lambda_{2,2}=\Lambda_2$ and $P_2=P_{2,1}=P_{2,2}$ naturally. For comparison purposes, we have included two benchmark schemes using Gaussian codes and shell codes. Moreover, we assume that the SIC in the benchmark schemes is always perfect regardless of blocklength. Note that in the scalar Gaussian broadcast channel, rate-splitting multiple access becomes superposition coding and SIC \cite[Remark 3]{RSMA_JSAC}. We stress that the perfect SIC assumption in the benchmark schemes is used for comparison purposes only. Whereas in practice, SIC introduces extra delay and complexity, and error propagation, which may not be suitable for URLLC services. In addition, one should be aware that when shell codes are used, the combination of inference and noise is neither shell codes nor Gaussian. Unlike Gaussian codes for which the dispersion can be computed by substituting the SINR into the dispersion function given in Section \ref{sec:preliminaries}, the dispersion of shell codes in the presence of interfering shell codes and Gaussian noise should follow \cite[Eq. (23)]{7605463}.
The derivation of second-order achievable rates for the benchmark schemes with perfect SIC follow \cite{9518265,9838392}. For shell codes, we optimistically assume that the each coded symbol is independent. Exact characterization would require the use of Berry-Esseen theorem for functions of random variables \cite[Th. 3]{7605463} since the coded symbols of shell codes are not independent. Because of this and with the perfect SIC assumption, the estimated achievable rates for the benchmark schemes are likely to be higher than the actual ones. Similar to \cite[Eq. (1)]{5452208}, the achievable rates of all schemes are approximated by their first two term, i.e., without the third-order terms.

\begin{figure}[t!]
	\centering
\includegraphics[width=\linewidth]{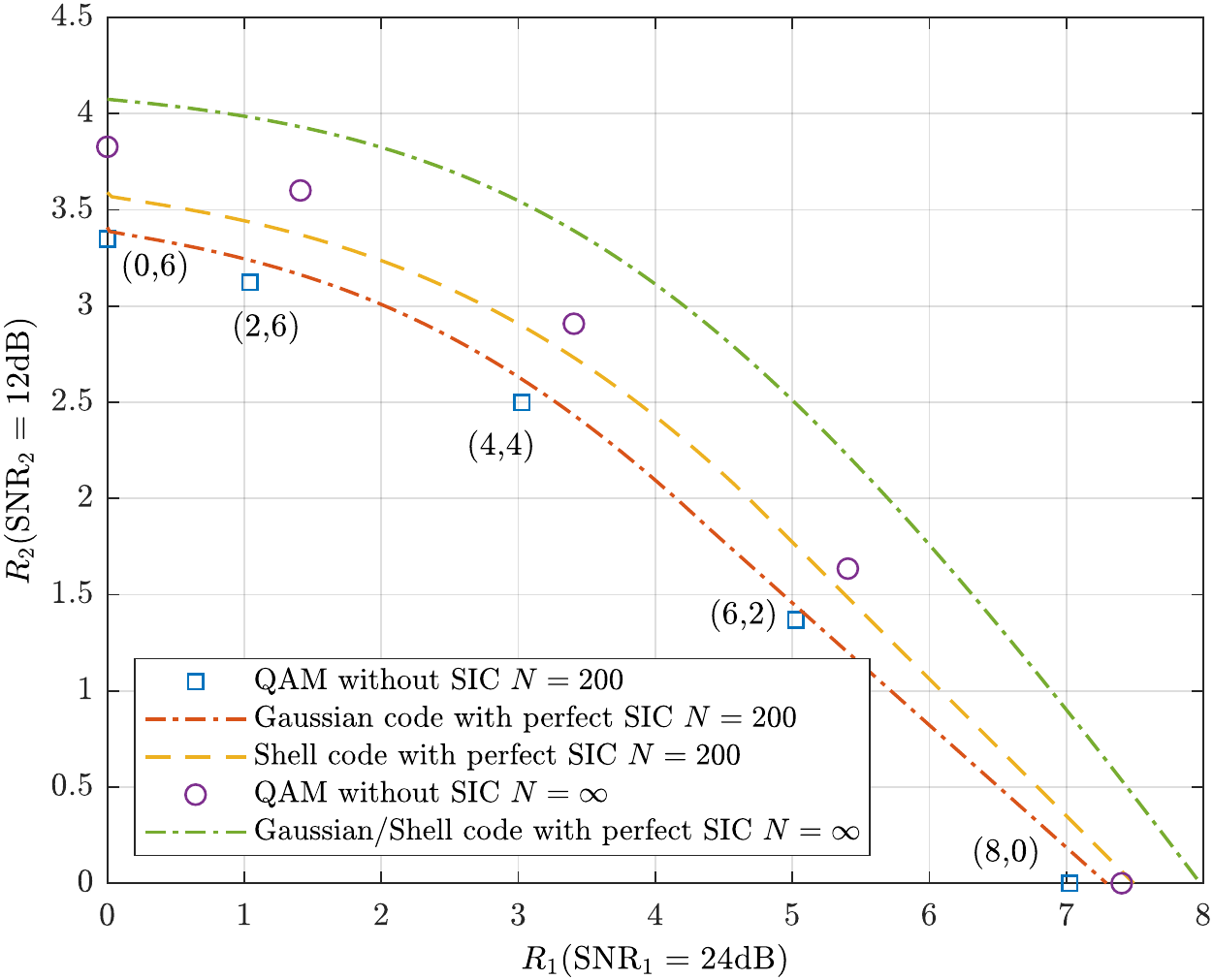}
\caption{Achievable rate pairs of the two-user case under homogeneous blocklength and error probability
constraints.}
\label{fig:N200}
\end{figure}

\begin{figure}[ht!]
	\centering
\includegraphics[width=\linewidth]{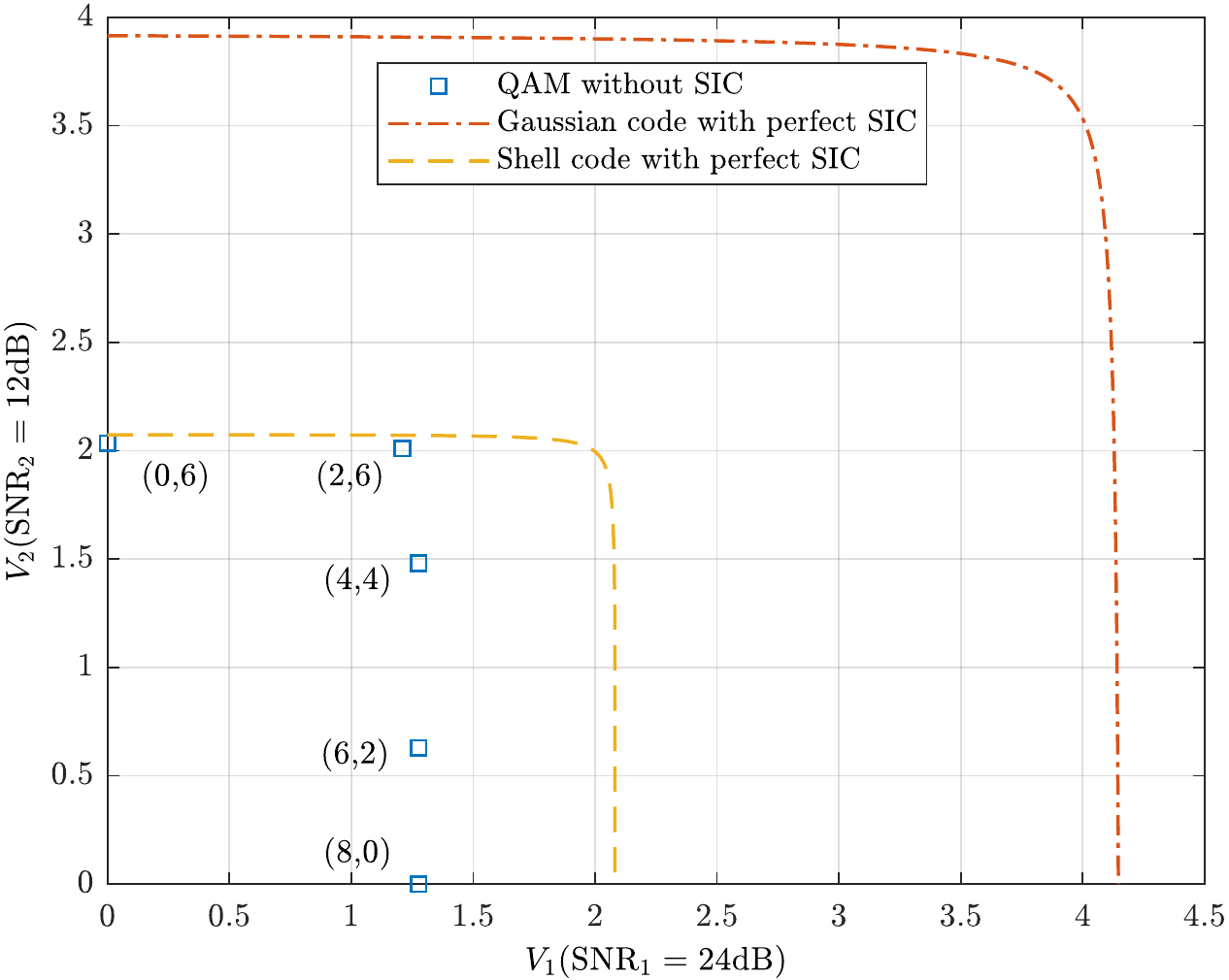}
\caption{Dispersion pairs of the two-user case under homogeneous blocklength and error probability
constraints.}
\label{fig:v12}
\end{figure}

Fig. \ref{fig:N200} shows the second-order achievable rate pairs of the proposed scheme with QAM and TIN and the benchmark schemes with perfect SIC under finite blocklength and infinite blocklength while Fig. \ref{fig:v12} shows the corresponding dispersion. In both figures, we have labeled the modulation orders $(m_1,m_2)$ for the proposed scheme. It can be seen that when the blocklength is small, the gap between the achievable rate of the proposed scheme and the benchmark scheme with Gaussian codes and perfect SIC is much smaller than that in the infinite blocklength regime. In fact, this can be explained by Fig. \ref{fig:v12} where the dispersion of QAM in the proposed scheme is shown to be much smaller than that of Gaussian codes and is no larger than that of shell codes. Since short blocklength and ultra-low target error probability are the main features of URLLC communication scenarios, the second-order term has a substantial impact on the achievable rate. Meanwhile, the achievable rate of the proposed scheme is close to the capacity region in the infinite blocklength case. Owing to the close-to-capacity first-order term, i.e., mutual information, and the smaller second-order term due to smaller channel dispersion, the proposed scheme can operate very close to the benchmark schemes with perfect SIC. Hence, the proposed scheme with QAM and TIN is attractive for supporting URLLC services.

\begin{figure}[t!]
	\centering
\includegraphics[width=\linewidth]{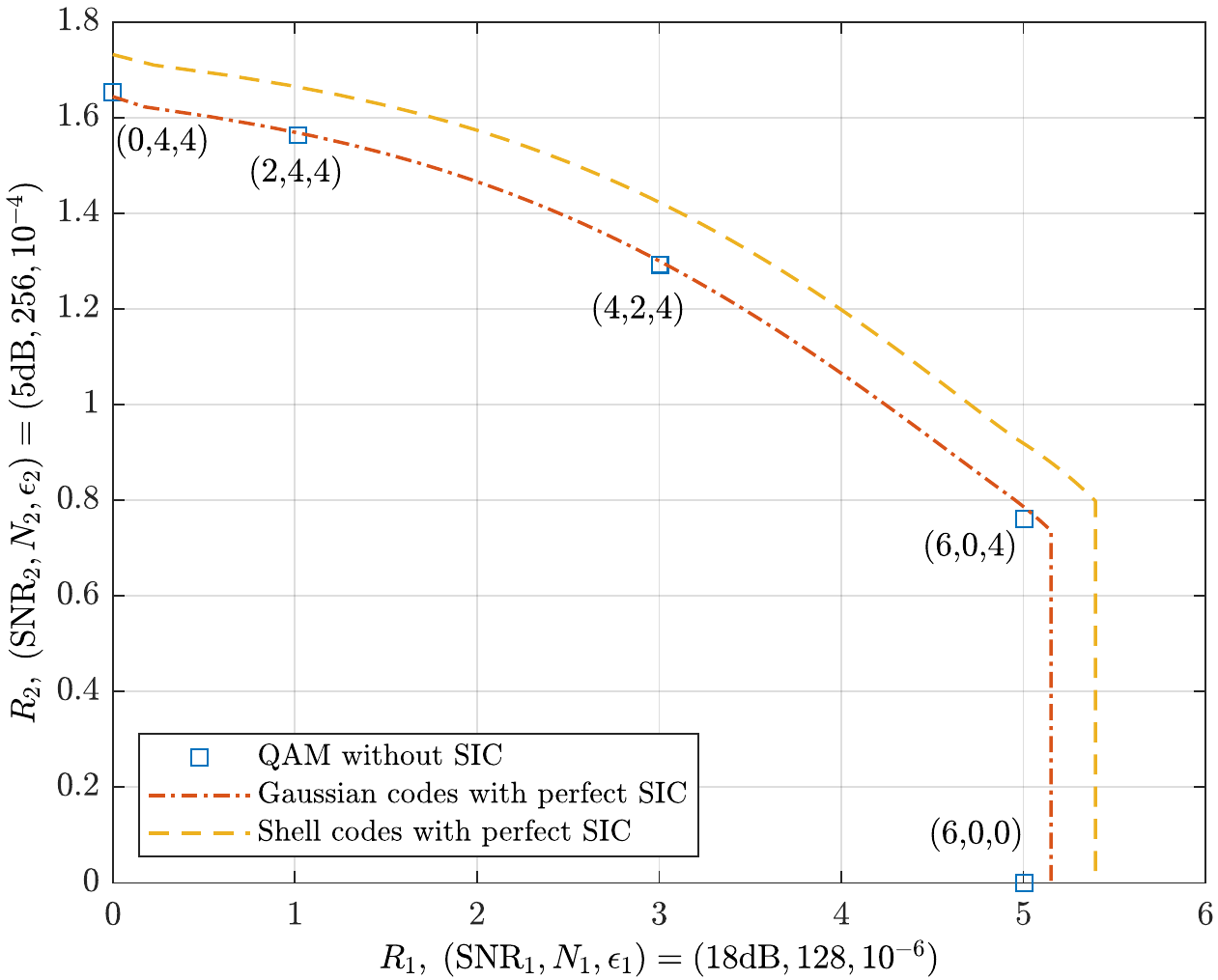}
\caption{Achievable rate pairs of the two-user case under heterogeneous blocklength and error probability
constraints.}
\label{fig:ar2}
\end{figure}

\begin{figure}[t!]
	\centering
\includegraphics[width=\linewidth]{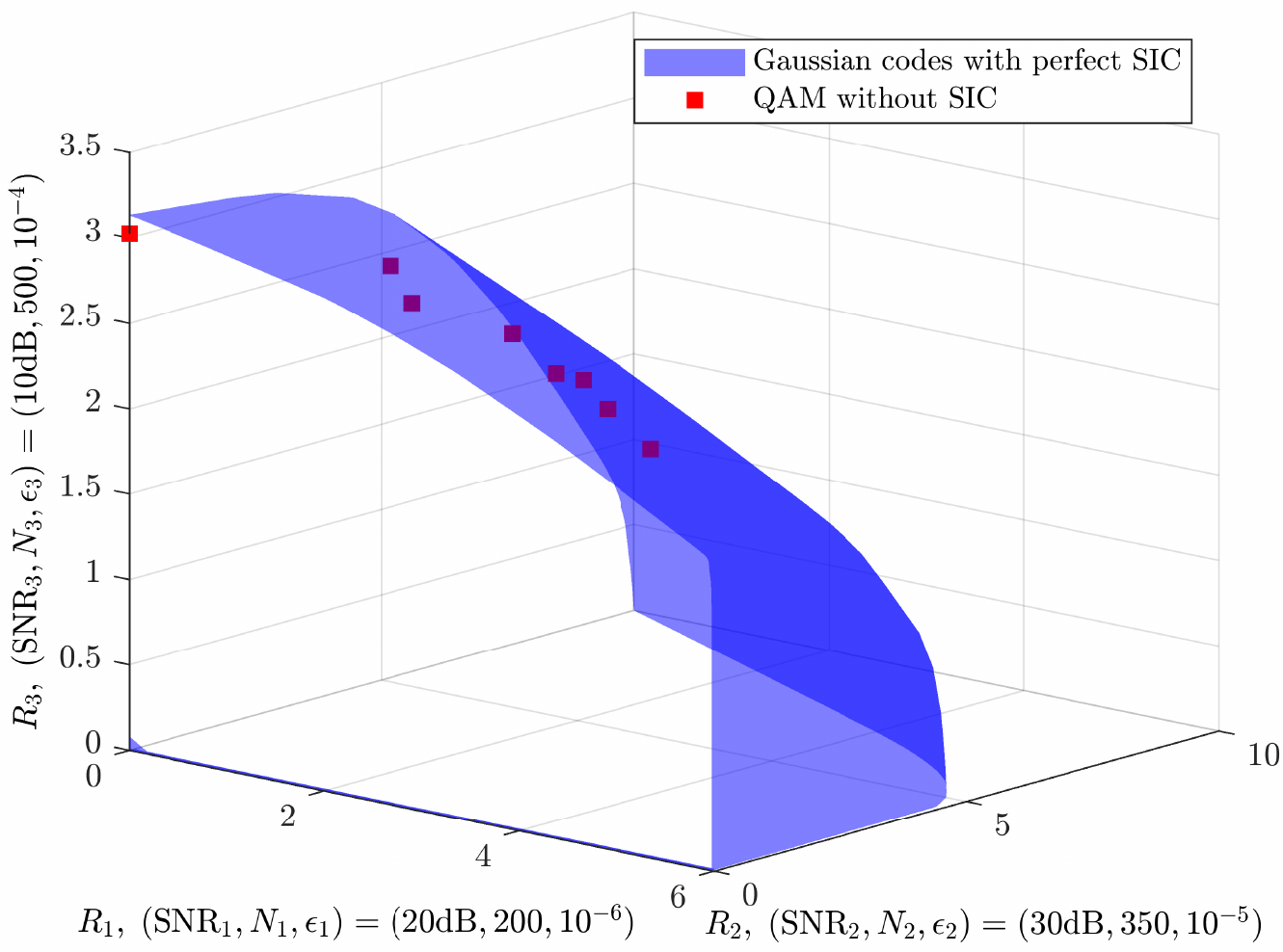}
\caption{Achievable rate triples of three-user case under heterogeneous blocklength and error probability
constraints.}
\label{fig:ar3}
\end{figure}

We then consider the heterogenous blocklength and error probability scenario, where $(\SNR_1,\SNR_2)=(18,5)$ in dB, $(N_1,N_2)=(128,256)$, and $(\epsilon_1,\epsilon_2)=(10^{-6},10^{-4})$. The achievable rate pairs of the proposed scheme and the aforementioned two benchmark schemes based on Gaussian and shell codes with perfect SIC are shown in Fig. \ref{fig:ar2}, where the modulation orders $(m_1,m_{2,1},m_{2,2})$ for the proposed scheme are labeled in the figure. Note that the proposed scheme uses the power assignment according to Section \ref{sec:2up} while the two benchmark schemes use brute-force search for $(P_1,P_{2,1},P_{2,2})$ to obtain their largest possible rate regions. Interestingly, the proposed scheme can still achieve rate pairs very close to those under Gaussian signaling and perfect SIC. Note that the behavior of channel dispersion is similar to that in Fig. \ref{fig:v12}, although it is not shown due to space limitation. This demonstrates that the proposed scheme is promising in supporting heterogeneous services. Notice that user 1's rate can remain to be the single-user rate while user 2's rate is increasing as shown in the bottom right corner of Fig. \ref{fig:ar2}. This is achieved by letting $P_{2,1}=0$ such that increasing $P_{2,2}$ does not affect user 1's rate. Apart from the two-user case, we also showcase the proposed scheme in the three-user case in Fig. \ref{fig:ar3}, where the channel parameters are given in the figure. For ease of exposition, we only show the largest rate region achieved by Gaussian codes with perfect SIC, which is based on exhaustive search of all combinations of sub-block power assignments. Similar observations can be made that the achievable rate triples of the proposed scheme with QAM and TIN are close to the benchmark scheme.

\subsection{Error Probability}\label{sec:err}

\begin{figure}[t!]
	\centering
\includegraphics[width=\linewidth]{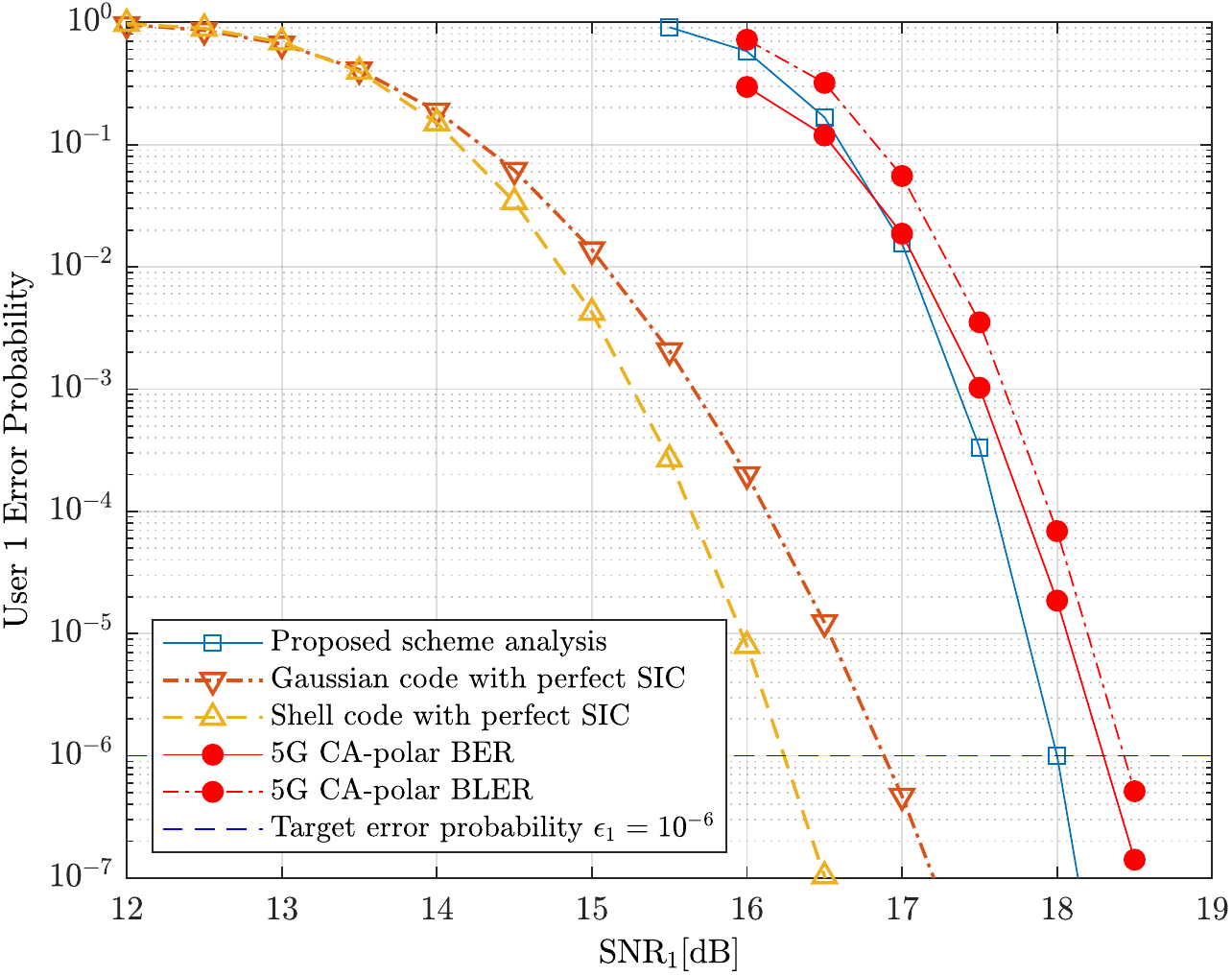}
\caption{Error probability of user 1.}
\label{fig:ber1}
\end{figure}

\begin{figure}[ht!]
	\centering
\includegraphics[width=\linewidth]{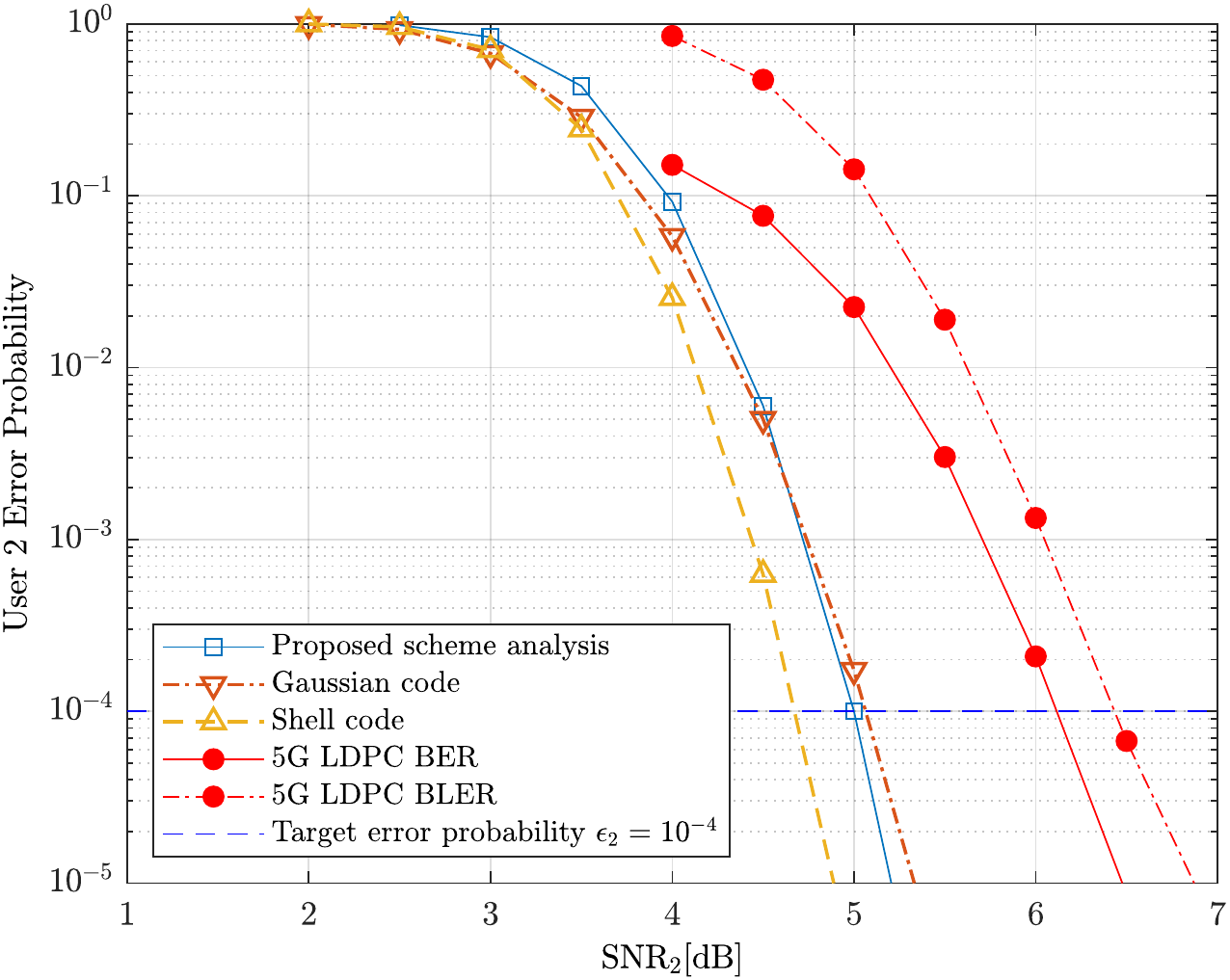}
\caption{Error probability of user 2.}
\label{fig:ber2}
\end{figure}

We build a practical set-up of the proposed scheme and evaluate the error performance for achieving a target rate pair by using off-the-shelf LDPC codes and polar codes. For illustrative purpose, we consider the same channel setting as for simulating Fig. \ref{fig:ar2}, where the proposed scheme with modulation orders $(m_1,m_{2,1},m_{2,2})=(2,4,4)$ achieves a rate pair of $(R_1,R_2) = (1.0174,1.5644)$ with $(\SNR_1,\SNR_2)=(18,5)$ in dB and error probability $(\epsilon_1,\epsilon_2)=(10^{-6},10^{-4})$. Since $(N_1,N_2)=(128,256)$, the channel codes for users 1 and 2 are with $(n_1,k_1)=(256,130)$ and $(n_2,k_2)=(1024,400)$, respectively. Specifically, user 1 employs a 5G standard CRC-aided polar (CA-polar) with an 11-bit CRC \cite{TS138212_v16p8} and adopts successive-cancellation list (SCL) decoding \cite{7055304} with list size 32. User 2 uses a 5G standard LDPC code from base graph 2 and with lifting size 40 \cite{TS138212_v16p8} and 976 bits punctured, and adopts layered belief propagation (BP) decoding \cite{1363033} with 50 maximum decoding iterations\footnote{In the conference version of this work \cite{ICCQiu23}, user 2 employs a 5G CA-polar code with SCL decoding, which exhibits a smaller gap to the analytical bound than that with the 5G LDPC code here.}. It is worth noting that this set-up can be regarded as serving heterogeneous devices. The bit error rate (BER) and block error rate (BLER) for users 1 and 2 are reported in Figs. \ref{fig:ber1}-\ref{fig:ber2}, respectively. We also include the upper bound of the average block error probability of the benchmark schemes with Gaussian codes and shell codes with perfect SIC as well as that of the proposed scheme with QAM and TIN in the same figure. The average block error probability upper bound is computed by rearranging the second-order achievable rates \eqref{eq:u1rate} and \eqref{eq:u2rate}. Note that all schemes achieve the same target rate pair and use the same power allocation.

It can be observed that the BER and BLER of user 1 are closer to the average block error probability upper bound of QAM at $10^{-6}$ than that for user 2 at $10^{-4}$. In fact, this behavior is similar to the single-user case \cite{8594709}, where CA-polar codes with short blocklength under SCL decoding perform closer to their respective finite blocklength error probability bound for binary phase-shift keying than LDPC codes with short-to-moderate blocklength under BP decoding. This implies that the proposed scheme allows the good performance of a code on the point-to-point AWGN channel to be \emph{carried over} to the considered multiuser channel under heterogeneous interference. It is also interesting to see that for user 2, the error probability upper bound for QAM slightly outperforms that of Gaussian signaling at $10^{-4}$ and below. This demonstrates that the proposed scheme with low-complexity TIN decoding is very promising at short blocklength. For user 1, the error probability upper bound for QAM is about 1 dB away from that of the Gaussian code at $10^{-6}$. This is because user 1 does not perform SIC in the proposed scheme while the benchmark schemes assume perfect SIC. Hence, the error performance of the corresponding coded systems also shows similar behavior. In summary, the proposed scheme using off-the-shelf codes can achieve satisfactory performance when compared to the error performance of shell codes with perfect SIC assumption. Meanwhile, the error probability upper bounds from \eqref{eq:u1rate} and \eqref{eq:u2rate} also serve as good performance indicators for practical coded modulation systems in the downlink BC model. To improve the error performance of the considered set-up, one can explicitly design codes and interleavers for the considered channel setting and use iterative detection and decoding \cite{CIT-019} at the cost of increased complexity. Furthermore, to approach the performance of shell codes, one can use multi-dimensional constellations \cite{8291591} with higher shaping gains.

\section{Conclusion}
In this paper, we have proposed a new coexistence scheme for simultaneously serving heterogeneous URLLC services by using discrete signaling and TIN. To effectively handle heterogeneous interference across received symbol sequences, we have divided the symbol block of each user into sub-blocks and designed the modulation and power for each sub-block. To characterize the behavior of practical coded modulations in the considered scenario, we have derived the second-order achievable rate under practical modulations and TIN with heterogeneous blocklength and error probability requirements. Simulation results have shown that under short blocklength constraints, the proposed scheme with QAM and TIN can operate very close to the benchmark schemes that assume perfect SIC with Gaussian signaling. This implies that practical coded modulations together with the low-complexity SUD are very promising for supporting downlink multiplexing of heterogeneous URLLC services with desired latency and reliability requirements while achieving near-optimal rates.

For future works, it would be interesting to extend the proposed discrete signaling and TIN to the uplink MAC where URLLC and other heterogeneous devices simultaneously communicate to the base station. Another worthwhile direction could be designing short channel codes with high shaping gain for the proposed scheme to further approach the performance of capacity-achieving signaling.

\appendices
\section{Useful Propositions and Lemmas}\label{APP1}
We first state a lemma regarding the power average over a sequence of i.i.d. symbols and the power of the normalized constellation.
\begin{lemma}\label{lem:power_constraint}
Consider a sequence of $N$ i.i.d. discrete random variables $X[1],\ldots,X[N]$, where $X[j]$ is uniformly distributed over a discrete constellation $\mathcal{X}$ and bounded $|X[j]|^2 \leq M<\infty$ for $j\in\{1,\ldots,N\}$. Moreover, we normalize the constellation points to satisfy the power constraint $\sum_{x\in\mathcal{X}}|x|^2=P$. Since $X[j]$ is i.i.d., we let $\sigma^2=\text{Var}(|X[j]|^2)$ for $j\in\{1,\ldots,N\}$. For any $\varepsilon>0$,
\begin{align}\label{eq:prop1}
\mathbb{P}\left[\frac{1}{N}\sum^N_{j=1}|X[j]|^2-P\geq\varepsilon\right]\leq e^{-\frac{N\varepsilon^2}{2(\sigma^2+M\varepsilon/3)}}.
\end{align}
\end{lemma}
\begin{IEEEproof}
Eq. \eqref{eq:prop1} is obtained by the Bernstein inequality \cite[Lemma A, Ch. 2.5.4]{GVK024353353}.
\end{IEEEproof}

We then introduce an upper bound on the decoding error probability, called the dependence testing bound. This bound allows us to obtain the same second-order terms of the achievable rate as using the random coding union bound \cite[Th. 16]{5452208} but with fewer steps.
\begin{proposition}[Th. 17 of \cite{5452208}]\label{RCU}
Consider a memoryless channel with the channel input and output $(X^{[n]},Y^{[n]})$ distributed as $P_{X^{[n]}}P_{Y^{[n]}|X^{[n]}}$. There exists an $(M,n,\epsilon)$ code $\mathcal{C}=\{X^{[n]}(1),\ldots,X^{[n]}(M)\}$ whose average error probability as a function of $n$ is upper bounded by
\begin{align}\label{eq:RCU}
\epsilon(n) \leq \E\left[2^{-\max\left\{0,i(X^{[n]};Y^{[n]})-\log\frac{M-1}{2}\right\}}\right],
\end{align}
\end{proposition}

\begin{lemma}\label{lem:err1}
The average error probability of \eqref{eq:RCU} in Proposition \ref{RCU} can be further bounded by
\begin{align}
\epsilon(n) \leq \mathbb{P}\left[ \frac{M-1}{2}2^{-i(X^{[n]};Y^{[n]})}>\frac{1}{n^s}\right]+\frac{1}{n^s},
\end{align}
for any $s\in(0,1]$.
\end{lemma}
\begin{IEEEproof}
Starting from \eqref{eq:RCU}, we have
\begin{align}
\epsilon(n)\leq&\E\left[2^{-\max\left\{0,i(X^{[n]};Y^{[n]})-\log\frac{M-1}{2}\right\}}\right] \nonumber\\
=&\E\left[\min\left\{1,\frac{M-1}{2}2^{-i(X^{[n]};Y^{[n]})}\right\}\right] \\
\leq& \mathbb{P}\left[ \frac{M-1}{2}2^{-i(X^{[n]};Y^{[n]})}> \frac{1}{n^s}\right]+\frac{1}{n^s},\label{eq:Emin_ineq}
\end{align}
where \eqref{eq:Emin_ineq} follows that for any $s\in(0,1]$ and nonnegative $X$, we have
\begin{align}
&\E[\min\{1,X\}] \nonumber\\
=& \int^{\infty}_{1}P_X(x)dx+\int^1_{\frac{1}{n^s}} xP_X(x) dx+ \int^{\frac{1}{n^s}}_{0} xP_X(x) dx \\
\leq & \int^{\infty}_{1}P_X(x)dx+\int^1_{\frac{1}{n^s}} P_X(x) dx+ \frac{1}{n^s}\int^{\frac{1}{n^s}}_{0} P_X(x) dx \\
=& \mathbb{P}\left[X > \frac{1}{n^s}\right]+\frac{1}{n^s}\cdot\mathbb{P}\left[X \leq \frac{1}{n^s}\right] 
\leq \mathbb{P}\left[X >\frac{1}{n^s}\right]+\frac{1}{n^s}.
\end{align}
This completes the proof.
\end{IEEEproof}

Next, we present the well-known Berry-Esseen central limit theorem that will be used for the derivation of second-order achievable rates.
\begin{proposition}[Th. 2, Ch. XVI-5 in \cite{Feller_book}]\label{CLT}
Let $U_j$, $j=1,\ldots,N$ be independent random variables with mean $\mu_j=\E[U_j]$, variance $\sigma^2_j=\text{Var}[U_j]>0$, and third absolute moment $t_j=\E[|U_j-\mu_j|^3]<\infty$. Then for any $-\infty<\lambda<\infty$, there exists a positive constant $C_0$ such that
\begin{align}\label{eq:CLT}
\left|\mathbb{P}\left[\frac{\sum^N_{j=1}(U_j-\mu_j)}{\sqrt{\sum^N_{j=1}\sigma^2_j}} \geq \lambda \right]-Q(\lambda)\right| \leq \frac{C_0\sum^N_{j=1}t_j}{\left(\sum^N_{j=1}\sigma^2_j\right)^\frac{3}{2}},
\end{align}
where $C_0$ has been refined to be 0.5600 in \cite{Shevtsova2010}.
\end{proposition}

\section{Proof of Achievability}
\subsection{Proof of Proposition \ref{prof:u1}}\label{app:u1}

We denote by $M_1$ the codebook size for user 1. Due to TIN decoding, we are able to use Lemma \ref{lem:err1} to upper bound the decoding error for user 1 as a function of $N_1$ as
\begin{align}
\epsilon_1(N_1)-&\frac{1}{(N_1)^s}  \leq \mathbb{P}\left[ \frac{M_1-1}{2}2^{-\sum_{j=1}^{N_1}i(X_1[j];Y_1[j])}> \frac{1}{(N_1)^s}\right] \label{eq:e1_upper_step1}\\
=&\Pp \left[\frac{\sum_{j=1}^{N_1}i(X_1[j];Y_1[j])-N_1I(X_1;Y_1)}{\sqrt{N_1V(X_1;Y_1)}}\right.\nonumber \\&<\left. \frac{\log \frac{M_1-1}{2} +s\log N_1-N_1I(X_1;Y_1)}{\sqrt{N_1 V(X_1;Y_1)}} \right], \label{eq:e1_upper}
\end{align}
where in \eqref{eq:e1_upper_step1} we have used \eqref{eq:u1id_basic} and in \eqref{eq:e1_upper} we have used \eqref{eq:NI1} and \eqref{eq:N1V1}. In order to use Proposition \ref{CLT} to obtain the second-order achievable rate, we first let $U_j=i(X_1[j];Y_1[j])$ for $j=1,\ldots,N_1$, which is i.i.d., and also let $\lambda = \frac{\log \frac{M_1-1}{2}+s\log N_1-N_1I(X_1;Y_1)}{\sqrt{N_1 V(X_1;Y_1)}}$. By substituting $\lambda$ and \eqref{eq:e1_upper} into \eqref{eq:CLT} and using the fact that $Q(\lambda)=1-Q(-\lambda)$, we have
\begin{align}\label{eq:CLT_u1}
\epsilon_1(N_1) \leq& Q\left( \frac{N_1I(X_1;Y_1)-\log \frac{M_1-1}{2} -s\log N_1}{\sqrt{N_1 V(X_1;Y_1)}}\right)\nonumber \\
&+ \frac{1}{(N_1)^s}+ \frac{B_1}{\sqrt{N_1}} \leq \epsilon_1,
\end{align}
where $B_1=\frac{C_0\E[|i(X_1;Y_1)-I(X_1;Y_1)|^3]}{(V(X_1;Y_1))^\frac{3}{2}}$ and the last inequality ensures that $\epsilon_1(N_1)$ in \eqref{eq:CLT_u1} is upper bounded by $\epsilon_1$ for all $N_1$. Then, one can solve $\log (M_1-1)$ from the last inequality of \eqref{eq:CLT_u1} and get
\begin{align}
&\log (M_1-1) \leq N_1I(X_1;Y_1)\nonumber \\
&-\sqrt{N_1 V(X_1;Y_1)}Q^{-1}\left(\epsilon_1 - \frac{1}{(N_1)^s}-\frac{B_1}{\sqrt{N_1}} \right) \nonumber \\
&+1-s\log N_1\label{eq:qinv1}\\
\Rightarrow& \log M_1 \leq \log (M_1-1)+1 \nonumber \\
&\leq N_1I(X_1;Y_1)-\sqrt{N_1 V(X_1;Y_1)}Q^{-1}\left(\epsilon_1\right)+O(\log N_1) \label{eq:taylor1},
\end{align}
where in \eqref{eq:qinv1} we have assumed that $N_1$ is large enough such that $(\epsilon_1 - \frac{1}{(N_1)^s}-\frac{B_1}{\sqrt{N_1}} ) \in (0,\frac{1}{2})$ as in \cite{5452208}, and in \eqref{eq:taylor1} we assume $M_1\geq2$ and choose $s\geq \frac{1}{2}$ and apply the first-order Taylor expansion to $Q^{-1}(.)$ such that $\sqrt{N_1 V(X_1;Y_1)}Q^{-1}(\epsilon_1 - \frac{1}{(N_1)^s}-\frac{B_1}{\sqrt{N_1}} )=\sqrt{N_1 V(X_1;Y_1)}Q^{-1}(\epsilon_1)+O(1)$. By dividing $N_1$ for both sides of \eqref{eq:taylor1}, we obtain \eqref{eq:u1rate}.

\subsection{Proof of Proposition \ref{prof:u2}}\label{app:u2}

We denote by $M_2$ the codebook size for user 2. Similar to user 1, by using Lemma \ref{lem:err1} with some manipulation, we upper bound the TIN decoding error probability for user 2 as a function of $N_2$ as
\begin{align}
&\epsilon_2(N_2)\leq\nonumber \\
&\Pp \left[\frac{\sum_{j=1}^{N_2}i(X_2[j];Y_2[j])-\sum^2_{i=1}(N_i-N_{i-1})I(X_{2,i};Y_{2,i})}{\sqrt{\sum^2_{i=1}(N_i-N_{i-1})V(X_{2,i};Y_{2,i})}}\right.\nonumber \\
&< \left.\frac{\log \frac{M_2-1}{2} +s\log N_2-\sum^2_{i=1}(N_i-N_{i-1})I(X_{2,i};Y_{2,i})}{\sqrt{\sum^2_{i=1}(N_i-N_{i-1})V(X_{2,i};Y_{2,i})}} \right]\nonumber \\
&+\frac{1}{(N_2)^s},\label{eq:e2_upper}
\end{align}
where in \eqref{eq:e2_upper} we have used \eqref{eq:NI2} and \eqref{eq:N2V2} and set $N_0=0$. In order to use Proposition \ref{CLT}, we let $U_j=i(X_2[j];Y_2[j])$ for $j=1,\ldots,N_2$. Note that $U_j$ for $j=1,\ldots,N_1$ are i.i.d. and that for $j=N_1+1,\ldots,N_2$ are i.i.d.. Let $\lambda = \frac{\log \frac{M_2-1}{2}+s\log N_2-\sum^2_{i=1}(N_i-N_{i-1})I(X_{2,i};Y_{2,i})}{\sqrt{\sum^2_{i=1}(N_i-N_{i-1})V(X_{2,i};Y_{2,i})}}$ and substitute it and \eqref{eq:e2_upper} into \eqref{eq:CLT}, we have
\begin{align}
&\epsilon_2(N_2)\leq\nonumber \\
 Q&\left( \frac{\sum^2_{i=1}(N_i-N_{i-1})I(X_{2,i};Y_{2,i})-\log \frac{M_2-1}{2} -s\log N_2}{\sqrt{\sum^2_{i=1}(N_i-N_{i-1})V(X_{2,i};Y_{2,i})}}\right) \nonumber \\
&+ \frac{1}{(N_2)^s}+\frac{B_2}{\sqrt{N_2}}\leq \epsilon_2 \label{eq:taylor_u2_initial}\\
\Rightarrow& \log M_2\leq \log(M_2-1)+1 \nonumber \\
&\leq\sum\nolimits^2_{i=1}(N_i-N_{i-1})I(X_{2,i};Y_{2,i})\nonumber \\
&-\sqrt{\sum\nolimits^2_{i=1}(N_i-N_{i-1})V(X_{2,i};Y_{2,i})}Q^{-1}\left(\epsilon_2\right)\nonumber \\
&+O(\log N_2)\label{eq:taylor_u2},
\end{align}
where in \eqref{eq:taylor_u2_initial} we have $B_2=\frac{C_0\sum^2_{i=1}\frac{N_i-N_{i-1}}{N_2}\mathbb{E}[|i(X_{2,i};Y_{2,i})-I(X_{2,i};Y_{2,i})|^3]}{(\sum^2_{i=1}\frac{N_i-N_{i-1}}{N_2}V(X_{2,i};Y_{2,i}))^\frac{3}{2}}$ and enforce $\epsilon_2(N_2)$ in \eqref{eq:e2_upper} to be upper bounded by $\epsilon_2$ for all $N_2$, in \eqref{eq:taylor_u2} we have assumed $M_2 \geq 2$ and used the first-order Taylor expansion of $Q^{-1}(.)$ about $\epsilon_2$, and choose $s \geq \frac{1}{2}$ similar to the approaches in Appendix \ref{app:u1}. Dividing both sides of \eqref{eq:taylor_u2} by $N_2$ leads to \eqref{eq:u2rate}.

\subsection{Proof of Proposition \ref{prop:uk}}\label{app:uk}

We denote by $M_k$ the codebook size for user $k$. By using Lemma \ref{lem:err1} with some manipulation, we upper bound the TIN decoding error probability for user $k$ as
\begin{align}
&\epsilon_k(N_k)\leq   \nonumber \\
&\Pp \left[\frac{\sum_{j=1}^{N_k}i(X_k[j];Y_k[j])-\sum^k_{j=1}(N_j-N_{j-1})I(X_{k,j};Y_{k,j})}{\sqrt{\sum^k_{j=1}(N_j-N_{j-1})V(X_{k,j};Y_{k,j})}}\right.\nonumber \\
&<\left. \frac{\log \frac{M_k-1}{2} +s\log N_k-\sum^k_{j=1}(N_j-N_{j-1})I(X_{k,j};Y_{k,j})}{\sqrt{\sum^k_{j=1}(N_j-N_{j-1})V(X_{k,j};Y_{k,j})}} \right]\nonumber \\
&+\frac{1}{(N_k)^s}\label{eq:ek_upper},
\end{align}
where we set $N_0=0$. To use Proposition \ref{CLT}, we let $U_j=i(X_k[j];Y_k[j])$ for $j=1,\ldots,N_k$. We know that $U_j$ is i.i.d. for $j=N_{j'-1}+1,\ldots,N_{j'}$, where $j'=1,\ldots,k$. Next, we let $\lambda = \frac{\log \frac{M_k-1}{2}+s\log N_k-\sum^k_{j=1}(N_j-N_{j-1})I(X_{k,j};Y_{k,j})}{\sqrt{\sum^k_{j=1}(N_j-N_{j-1})V(X_{k,j};Y_{k,j})}}$ and substitute it and \eqref{eq:ek_upper} into \eqref{eq:CLT}, we have
\begin{align}
&\epsilon_k(N_k)   \leq \nonumber \\
 Q&\left( \frac{\sum^k_{j=1}(N_j-N_{j-1})I(X_{k,j};Y_{k,j}) - \log \frac{M_k-1}{2} - s\log N_k}{\sqrt{\sum^k_{j=1}(N_j-N_{j-1})V(X_{k,j};Y_{k,j})}}\right)\nonumber \\
&+  \frac{1}{(N_k)^s}+\frac{B_k}{\sqrt{N_k}}\leq \epsilon_k\label{eq:taylor_uk_initial} \\
\Rightarrow& \log M_k \leq \log (M_k-1)+1 \nonumber \\
&\leq \sum\nolimits^k_{j=1}(N_j-N_{j-1})I(X_{k,j};Y_{k,j})\nonumber \\
&-\sqrt{\sum\nolimits^k_{j=1}(N_j-N_{j-1})V(X_{k,j};Y_{k,j})}Q^{-1}\left(\epsilon_k\right)\nonumber \\
&+O(\log N_k),\label{eq:taylor_uk}
\end{align}
where in \eqref{eq:taylor_uk_initial} we have $B_k=\frac{C_0\sum^k_{j=1}\frac{N_j-N_{j-1}}{N_k}\mathbb{E}[|i(X_{k,j};Y_{k,j})-I(X_{k,j};Y_{k,j})|^3]}{(\sum^k_{j=1}\frac{N_j-N_{j-1}}{N_k}V(X_{k,j};Y_{k,j}))^\frac{3}{2}}$ and enforce $\epsilon_k(N_k)$ in \eqref{eq:ek_upper} to be upper bounded by $\epsilon_k$ for all $N_k$, in \eqref{eq:taylor_uk} we have assumed $M_k\geq2$ and used the first-order Taylor expansion of $Q^{-1}(.)$ about $\epsilon_k$ and choose $s \geq \frac{1}{2}$ as in Appendix \ref{app:u1}. Dividing both sides of \eqref{eq:taylor_uk} by $N_k$ gives \eqref{eq:uk_rate}.

\bibliographystyle{IEEEtran}
\bibliography{MinQiu}

\begin{thebibliography}{10}
\providecommand{\url}[1]{#1}
\csname url@samestyle\endcsname
\providecommand{\newblock}{\relax}
\providecommand{\bibinfo}[2]{#2}
\providecommand{\BIBentrySTDinterwordspacing}{\spaceskip=0pt\relax}
\providecommand{\BIBentryALTinterwordstretchfactor}{4}
\providecommand{\BIBentryALTinterwordspacing}{\spaceskip=\fontdimen2\font plus
\BIBentryALTinterwordstretchfactor\fontdimen3\font minus
  \fontdimen4\font\relax}
\providecommand{\BIBforeignlanguage}[2]{{%
\expandafter\ifx\csname l@#1\endcsname\relax
\typeout{** WARNING: IEEEtran.bst: No hyphenation pattern has been}%
\typeout{** loaded for the language `#1'. Using the pattern for}%
\typeout{** the default language instead.}%
\else
\language=\csname l@#1\endcsname
\fi
#2}}
\providecommand{\BIBdecl}{\relax}
\BIBdecl

\bibitem{ICCQiu23}
M.~Qiu, Y.-C. Huang, and J.~Yuan, ``Downlink transmission under heterogeneous
  blocklength constraints: Discrete signaling with single-user decoding,'' in
  \emph{Proc. IEEE Int. Conf. Commun. (ICC)}, 2023, accepted.

\bibitem{TR38.913}
3GPP, ``Study on scenarios and requirements for next generation access
  technologies,'' {3rd Generation Partnership Project (3GPP)}, TR 38.913
  V17.0.0, May 2022.

\bibitem{7945856}
C.~She, C.~Yang, and T.~Q.~S. Quek, ``Radio resource management for
  ultra-reliable and low-latency communications,'' \emph{IEEE Commun. Mag.},
  vol.~55, no.~6, pp. 72--78, Jun. 2017.

\bibitem{8469808}
H.~Chen, R.~Abbas, P.~Cheng, M.~Shirvanimoghaddam, W.~Hardjawana, W.~Bao,
  Y.~Li, and B.~Vucetic, ``Ultra-reliable low latency cellular networks: Use
  cases, challenges and approaches,'' \emph{IEEE Commun. Mag.}, vol.~56,
  no.~12, pp. 119--125, Dec. 2018.

\bibitem{8594709}
M.~Shirvanimoghaddam, M.~S. Mohammadi, R.~Abbas, A.~Minja, C.~Yue, B.~Matuz,
  G.~Han, Z.~Lin, W.~Liu, Y.~Li, S.~Johnson, and B.~Vucetic, ``Short
  block-length codes for ultra-reliable low latency communications,''
  \emph{IEEE Commun. Mag.}, vol.~57, no.~2, pp. 130--137, Feb. 2019.

\bibitem{8004168}
H.~Zhang, N.~Liu, X.~Chu, K.~Long, A.-H. Aghvami, and V.~C.~M. Leung, ``Network
  slicing based 5{G} and future mobile networks: Mobility, resource management,
  and challenges,'' \emph{IEEE Commun. Mag.}, vol.~55, no.~8, pp. 138--145,
  2017.

\bibitem{TR38.802}
3GPP, ``Study on new radio ({NR}) access technology physical layer aspects,''
  {3rd Generation Partnership Project (3GPP)}, TR 38.802 V14.2.0, Sept. 2017.

\bibitem{8403963}
H.~Ji, S.~Park, J.~Yeo, Y.~Kim, J.~Lee, and B.~Shim, ``Ultra-reliable and
  low-latency communications in 5{G} downlink: Physical layer aspects,''
  \emph{IEEE Wireless Commun.}, vol.~25, no.~3, pp. 124--130, 2018.

\bibitem{8476595}
P.~Popovski, K.~F. Trillingsgaard, O.~Simeone, and G.~Durisi, ``5{G} wireless
  network slicing for e{MBB}, {URLLC}, and m{MTC}: A communication-theoretic
  view,'' \emph{IEEE Access}, vol.~6, pp. 55\,765--55\,779, 2018.

\bibitem{8647460}
R.~Kassab, O.~Simeone, and P.~Popovski, ``Coexistence of {URLLC} and e{MBB}
  services in the {C-RAN} uplink: An information-theoretic study,'' in
  \emph{Proc. IEEE Globecom}, 2018, pp. 1--6.

\bibitem{21p915}
3GPP, ``Technical specification group services and system aspects,'' {3rd
  Generation Partnership Project (3GPP)}, Tech. Spec., {TS 21.915 V1.1.0}, Mar.
  2019.

\bibitem{9562192}
O.~Dizdar, Y.~Mao, Y.~Xu, P.~Zhu, and B.~Clerckx, ``Rate-splitting multiple
  access for enhanced {URLLC} and e{MBB} in 6{G},'' in \emph{Proc. 17th Int.
  Symp. Wireless Commun. Syst. (ISWCS)}, Sep. 2021, pp. 1--6.

\bibitem{9831059}
F.~Saggese, M.~Moretti, and P.~Popovski, ``Power minimization of downlink
  spectrum slicing for e{MBB} and {URLLC} users,'' \emph{IEEE Trans. Wireless
  Commun.}, vol.~21, no.~12, pp. 11\,051--11\,065, 2022.

\bibitem{Cover:2006:EIT:1146355}
T.~M. Cover and J.~A. Thomas, \emph{Elements of Information Theory}.\hskip 1em
  plus 0.5em minus 0.4em\relax New York, NY, USA: Wiley-Interscience, 2006.

\bibitem{tse_book}
D.~Tse and P.~Viswanath, \emph{Fundamentals of Wireless Communication}.\hskip
  1em plus 0.5em minus 0.4em\relax New York, NY, USA: Cambridge University
  Press, 2005.

\bibitem{Ding17J}
Z.~Ding, X.~Lei, G.~K. Karagiannidis, R.~Schober, J.~Yuan, and V.~Bhargava, ``A
  survey on non-orthogonal multiple access for 5{G} networks: Research
  challenges and future trends,'' \emph{IEEE J. Sel. Areas Commun.}, vol.~35,
  no.~10, pp. 2181--2195, Oct. 2017.

\bibitem{8876877}
Z.~Wei, L.~Yang, D.~W.~K. Ng, J.~Yuan, and L.~Hanzo, ``On the performance gain
  of {NOMA} over {OMA} in uplink communication systems,'' \emph{IEEE Trans.
  Commun.}, vol.~68, no.~1, pp. 536--568, Jan. 2020.

\bibitem{9693417}
Y.~Liu, S.~Zhang, X.~Mu, Z.~Ding, R.~Schober, N.~Al-Dhahir, E.~Hossain, and
  X.~Shen, ``Evolution of {NOMA} toward next generation multiple access
  ({NGMA}) for 6{G},'' \emph{IEEE J. Sel. Areas Commun.}, vol.~40, no.~4, pp.
  1037--1071, Apr. 2022.

\bibitem{9831440}
Y.~Mao, O.~Dizdar, B.~Clerckx, R.~Schober, P.~Popovski, and H.~V. Poor,
  ``Rate-splitting multiple access: Fundamentals, survey, and future research
  trends,'' \emph{IEEE Commun. Surveys Tuts.}, vol.~24, no.~4, pp. 2073--2126,
  Fourthquarter 2022.

\bibitem{RSMA_JSAC}
\BIBentryALTinterwordspacing
B.~Clerckx, Y.~Mao, E.~A. Jorswieck, J.~Yuan, D.~J. Love, E.~Erkip, and
  D.~Niyato, ``A primer on rate-splitting multiple access: Tutorial, myths, and
  frequently asked questions,'' \emph{IEEE J. Sel. Areas Commun.}, 2023, early
  access. [Online]. Available:
  \url{https://ieeexplore.ieee.org/document/10038476}
\BIBentrySTDinterwordspacing

\bibitem{5452208}
Y.~Polyanskiy, H.~V. Poor, and S.~Verdu, ``Channel coding rate in the finite
  blocklength regime,'' \emph{IEEE Trans. Inf. Theory}, vol.~56, no.~5, pp.
  2307--2359, May 2010.

\bibitem{7529226}
G.~Durisi, T.~Koch, and P.~Popovski, ``Toward massive, ultrareliable, and
  low-latency wireless communication with short packets,'' \emph{Proc. IEEE},
  vol. 104, no.~9, pp. 1711--1726, Sep. 2016.

\bibitem{8277977}
A.~Ünsal and J.-M. Gorce, ``The dispersion of superposition coding for
  {Gaussian} broadcast channels,'' in \emph{IEEE Inf. Theory Workshop (ITW)},
  2017, pp. 414--418.

\bibitem{8345745}
X.~Sun, S.~Yan, N.~Yang, Z.~Ding, C.~Shen, and Z.~Zhong, ``Short-packet
  downlink transmission with non-orthogonal multiple access,'' \emph{IEEE
  Trans. Wireless Commun.}, vol.~17, no.~7, pp. 4550--4564, Jul. 2018.

\bibitem{8933345}
H.~Ren, C.~Pan, Y.~Deng, M.~Elkashlan, and A.~Nallanathan, ``Joint power and
  blocklength optimization for {URLLC} in a factory automation scenario,''
  \emph{IEEE Trans. Wireless Commun.}, vol.~19, no.~3, pp. 1786--1801, Mar.
  2020.

\bibitem{8909370}
Y.~Xu, C.~Shen, T.-H. Chang, S.-C. Lin, Y.~Zhao, and G.~Zhu, ``Transmission
  energy minimization for heterogeneous low-latency {NOMA} downlink,''
  \emph{IEEE Trans. Wireless Commun.}, vol.~19, no.~2, pp. 1054--1069, Feb.
  2020.

\bibitem{9518265}
P.-H. Lin, S.-C. Lin, and E.~A. Jorswieck, ``Early decoding for {G}aussian
  broadcast channels with heterogeneous blocklength constraints,'' in
  \emph{Proc. IEEE Int. Symp. Inf. Theory (ISIT)}, 2021, pp. 3243--3248.

\bibitem{9838392}
P.-H. Lin, S.-C. Lin, P.-W. Chen, M.~Mross, and E.~A. Jorswieck, ``Rate region
  of {G}aussian broadcast channels with heterogeneous blocklength
  constraints,'' in \emph{Proc. IEEE Int. Conf. Commun. (ICC)}, May 2022, pp.
  2144--2150.

\bibitem{6875095}
C.~Sahin, L.~Liu, and E.~Perrins, ``Early decoding for transmission over finite
  transport blocks,'' in \emph{Proc. IEEE Int. Symp. Inf. Theory (ISIT)}, 2014,
  pp. 1558--1562.

\bibitem{6767457}
C.~E. Shannon, ``Probability of error for optimal codes in a {G}aussian
  channel,'' \emph{Bell Syst. Tech. J.}, vol.~38, no.~3, pp. 611--656, May
  1959.

\bibitem{TS138212_v16p8}
3GPP, ``5{G};{NR}; {M}ultiplexing and channel coding,'' {3rd Generation
  Partnership Project (3GPP)}, TR 38.212, Jan. 2022.

\bibitem{8291591}
M.~Qiu, Y.-C. Huang, S.-L. Shieh, and J.~Yuan, ``A lattice-partition framework
  of downlink non-orthogonal multiple access without {SIC},'' \emph{IEEE Trans.
  Commun.}, vol.~66, no.~6, pp. 2532 -- 2546, Jun. 2018.

\bibitem{9535131}
M.~Qiu, Y.-C. Huang, and J.~Yuan, ``Discrete signaling and treating
  interference as noise for the {G}aussian interference channel,'' \emph{IEEE
  Trans. Inf. Theory}, vol.~67, no.~11, pp. 7253--7284, Nov. 2021.

\bibitem{Feller_book}
W.~Feller, \emph{An Introduction to Probability Theory and Its
  Applications}.\hskip 1em plus 0.5em minus 0.4em\relax New York, NY, USA:
  Wiley, 1971, vol.~II.

\bibitem{8108239}
G.~Geraci, D.~Fang, and H.~Claussen, ``A new method of {MIMO}-based
  non-orthogonal multiuser downlink transmission,'' in \emph{Proc. IEEE VTC
  Spring}, Jun. 2017, pp. 1--5.

\bibitem{GamalKim11}
A.~E. Gamal and Y.-H. Kim, \emph{Network Information Theory}.\hskip 1em plus
  0.5em minus 0.4em\relax Cambridge,U.K.: Cambridge Univ. Press, 2011.

\bibitem{7056434}
V.~Y.~F. Tan and M.~Tomamichel, ``The third-order term in the normal
  approximation for the {AWGN} channel,'' \emph{IEEE Trans. Inf. Theory},
  vol.~61, no.~5, pp. 2430--2438, May 2015.

\bibitem{7605463}
J.~Scarlett, V.~Y.~F. Tan, and G.~Durisi, ``The dispersion of nearest-neighbor
  decoding for additive non-{G}aussian channels,'' \emph{IEEE Trans. Inf.
  Theory}, vol.~63, no.~1, pp. 81--92, Jan. 2017.

\bibitem{8066336}
M.~Qiu, L.~Yang, Y.~Xie, and J.~Yuan, ``On the design of multi-dimensional
  irregular repeat-accumulate lattice codes,'' \emph{IEEE Trans. Commun.},
  vol.~66, no.~2, pp. 478--492, Feb. 2018.

\bibitem{CIT-019}
A.~G. i~Fàbregas, A.~Martinez, and G.~Caire, ``Bit-interleaved coded
  modulation,'' \emph{Found. Trends Commun. Inf. Theory}, vol.~5, no. 1–2,
  pp. 1--153, 2008.

\bibitem{1057022}
R.~Gallager, ``A perspective on multiaccess channels,'' \emph{IEEE Trans. Inf.
  Theory}, vol.~31, no.~2, pp. 124--142, Mar. 1985.

\bibitem{7055304}
I.~Tal and A.~Vardy, ``List decoding of polar codes,'' \emph{IEEE Trans. Inf.
  Theory}, vol.~61, no.~5, pp. 2213--2226, May 2015.

\bibitem{1363033}
D.~Hocevar, ``A reduced complexity decoder architecture via layered decoding of
  {LDPC} codes,'' in \emph{Proc. IEEE Workshop Signal Process. Syst. (SIPS)},
  Oct. 2004, pp. 107--112.

\bibitem{GVK024353353}
R.~Serfling, \emph{Approximation theorems of mathematical statistics}.\hskip
  1em plus 0.5em minus 0.4em\relax New York, NY, USA: Wiley, 1980.

\bibitem{Shevtsova2010}
I.~G. Shevtsova, ``An improvement of convergence rate estimates in the
  {L}yapunov theorem,'' \emph{Dokl. Math.}, vol.~82, no.~3, pp. 862--864, Dec.
  2010.

\end{thebibliography}

\end{document}